\documentclass[letterpaper, 10 pt, conference]{ieeeconf}  
\usepackage{amsmath} 
\usepackage{amssymb}  
\usepackage{graphicx}
\usepackage{amsmath}
\usepackage{epstopdf}
\usepackage{mathtools}
\usepackage{dsfont}
\usepackage{url}
\usepackage[ruled,norelsize]{algorithm2e}

\makeatletter
\newcommand{\removelatexerror}{\let\@latex@error\@gobble}
\makeatother

\newcommand{\tKinf}{\mathcal{K}_{\infty}}

\newcommand{\tKN}{\mathcal{K}_{N}}         
\DeclareMathOperator{\Succ}{Succ}
\DeclareMathOperator{\Pre}{Pre}
\DeclareMathOperator{\Int}{Int}
\newcommand{\beq}{\begin{equation}}
\newcommand{\eeq}{\end{equation}}
\newcommand{\rr}{{\mathbb R}}
\newcommand{\zz}{{\mathbb Z}}
\newcommand{\cvd}{\hfill$\blacksquare$\vspace{0em}\par\noindent}

\newcounter{algorithmctr}[section]
\renewcommand{\thealgorithmctr}{\thesection.\arabic{algorithmctr}}
   {\refstepcounter{algorithmctr}\begin{list}{}{%
       \setlength{\rightmargin}{0\linewidth}%
       \setlength{\leftmargin}{.05\linewidth}
        \setlength{\itemsep}{1pt}
  \setlength{\parskip}{0pt}
  \setlength{\parsep}{0pt}}%
       \rmfamily\small
       \item[]{\setlength{\parskip}{0ex}\hrulefill\par%
        \nopagebreak{\bfseries\textsf{Algorithm \thealgorithmctr~}}}}%
   {{\setlength{\parskip}{-1ex}\nopagebreak\par\hrulefill} \end{list}}
\IEEEoverridecommandlockouts
\newtheorem{assumption}{Assumption}
\newtheorem{theorem}{Theorem}
\newtheorem{remark}{Remark}

\newtheorem{definition}{Definition}

\usepackage{xcolor}
\usepackage[noadjust]{cite} 
\usepackage{balance}

\title{\LARGE \bf
Learning Model Predictive Control for Iterative Tasks. A Data-Driven Control Framework. }

\author{Ugo Rosolia and Francesco Borrelli
\thanks{Ugo Rosolia and Francesco Borrelli are with the Department of Mechanical Engineering, University of California at Berkeley ,
        Berkeley, CA 94701, USA
        {\tt\small \{ugo.rosolia, fborrelli\} $@$ berkeley.edu}}%
}

\begin{document}

\maketitle
\thispagestyle{empty}
\pagestyle{empty}

\begin{abstract}

A Learning Model Predictive Controller (LMPC) for iterative tasks is presented.
The controller is reference-free and is able to improve its performance by learning from previous iterations. A safe set and a terminal cost function are used in order to guarantee recursive feasibility and non-decreasing performance at each iteration.
The paper presents the control design approach, and shows how to recursively construct terminal set and terminal cost from state and input trajectories of previous iterations.
Simulation results show the effectiveness of the proposed control logic.
\end{abstract}

\section{INTRODUCTION}
Control systems autonomously performing a repetitive task have been extensively studied in the literature \cite{c3,c6,c7,c8,c21,c22}. One task execution is often referred to as ``iteration" or ``trial". Iterative Learning Control (ILC) is a control strategy that allows learning from previous iterations to improve its closed-loop tracking performance. In ILC, at each iteration, the system starts from the same initial condition and the controller objective is to track a given reference, rejecting periodic disturbances \cite{c3,c7}. The main advantage of ILC is that information from previous iterations are incorporated in the problem formulation at the next iteration and are used to improve the system performance.

The possibility of combining Model Predictive Control (MPC) with ILC has been explored in \cite{c34}, where the authors proposed a Model-based Predictive Control for Batch processes, called Batch-MPC (BMPC). The BMPC is based on a time-varying MIMO system that has a dynamic memory of past batches tracking error. The effectiveness of this approach has been shown through experimental results on a nonlinear system \cite{c34}, and in \cite{c33} the authors proved that the tracking error of the BMPC converges to zero as the number of iterations increases. In~\cite{c35} a model-based iterative learning control has been proposed. The authors incorporated the tracking error of the previous iterations in the control law and used an observer to deal with stochastic disturbances and noises. Also in this case, the authors showed that the tracking error asymptotically converges to zero. The works in \cite{c6, c4, liu2013nonlinear} also use MPC for repetitive tasks. In \cite{c6}, the authors successfully achieve zero tracking error using a  MPC which uses measurements from previous iterations to modify the cost function. In~\cite{c4} the authors use the trajectories of previous iterations to linearize the model used in the MPC algorithm. The authors proved zero steady-state tracking error in presence of model mismatch. In \cite{liu2013nonlinear}, a nonlinear MPC based on iterative learning control is proposed. There, a MPC is designed for disturbance rejection and the ILC is designed to minimize errors occurring at each iteration. The authors proved that the steady state tracking error converges to zero as the iteration index goes to infinity.
In all aforementioned papers the control goal is to minimize a tracking error under the presence of disturbances. The reference signal is known in advance and does not change at each iteration.

In this paper we are interested in repetitive tasks where the reference trajectory it is not known. In general, a reference trajectory that maximize the performance over an infinite horizon may be challenging to compute for a system with complex nonlinear dynamics or with parameter uncertainty. These systems include race and rally cars where the environment and the dynamics are complex and not perfectly known \cite{c17,c18}, or bipedal locomotion with exoskeletons where the human input is unknown apriori and can change at each iteration \cite{c19,c20}.

Our objective is to design a reference-free iterative control strategy able to learn from previous iterations. At each iteration the cost associated with  the closed-loop trajectory shall not increase and  state and input constraints shall be satisfied. Nonlinear MPC control is an appealing technique to tackle this problem for its ability to handle state and inputs constraints while minimizing a finite-time predicted cost \cite{c11}. However, the receding horizon nature can lead to infeasibility and it does not guaranty improved  performance at each iteration \cite{c12}.

The contribution of this paper is threefold. First we present
a novel reference-free learning MPC design for an iterative control task. At each iteration,  the initial condition, the constraints and the objective
function do not change. The $j$-th iteration  cost is defined as the objective function evaluated for the $j$-th closed loop system trajectory.
Second, we show how to design a terminal safe set and a terminal cost function
in order to guarantee that \emph{(i):} the $j$-th iteration  cost does not increase compared to the ($j{\text-}1$)-th iteration  cost (non-increasing cost at each iteration), \emph{(ii):} state and input constraints are satisfied at iteration $j$ if they were satisfied at iteration  $j{\text-}1$ (recursive feasibility), \emph{(iii):} the  closed-loop equilibrium is asymptotically stable.
Third, we assume that the system converges to a steady state trajectory as  the number of iteration $j$ goes to infinity and we prove the optimality of such trajectory for convex problems.

This paper is organized as follows: in Section II we formally define an iterative task and its $j$-th iteration cost. The control strategy is illustrated in Section III, where we show the recursive feasibility and stability of the control logic and prove the convergence properties.
Finally, in Section IV and V, we test the proposed control logic on an infinite horizon linear quadratic regulator with constraints and on a minimum time Dubins car problem. Section VI and VII provide final remarks.

\section{PROBLEM DEFINITION}

Consider the discrete time system
\begin{equation}\label{eq:system}
x_{t+1}=f(x_t,u_t),
\end{equation}
where $x\in \rr^n$ and $u\in\rr^{m}$ are the system state and input,
respectively. We assume that $f(\cdot,\cdot)$ is continuous and
 that state and inputs are subject to  the constraints
\begin{equation}
x_t \in \mathcal{X},\ u_t\in \mathcal{U},\ \forall t \geq 0.
\label{eq:inv_constraints}
\end{equation}

At the $j$-th iteration the vectors
\begin{subequations}\label{eq:sequence}
\begin{align}
 {\bf{u}}^j ~ = ~ [u_0^j,~u_1^j,~...,~u_t^j,~...], \label{eq:sequenceU} \\
 {\bf{x}}^j ~ = ~ [x_0^j,~x_1^j,~...,~x_t^j,~...], \label{eq:sequenceX}
\end{align}
\end{subequations}
collect the inputs applied to system~(\ref{eq:system}) and the corresponding
state evolution. In (\ref{eq:sequence}), $x_t^j$ and $u_t^j$ denote the system state and the control input at time $t$ of the $j$-th iteration. We assume that at each $j$-th iteration the closed loop trajectories start from the same initial state,
\begin{equation}\label{eq:Initial_state}
\begin{aligned}
 x_0^j ~ = x_S, ~\forall j \geq 0.\\
\end{aligned}
\end{equation}

The goal is to design a controller which solves the following infinite horizon optimal control problem at each iteration:
\begin{subequations}\label{eq:ConstraintsInf}
\begin{align}
J_{0\rightarrow \infty}^*(x_S)&=\min_{u_0,u_1,\ldots} \sum\limits_{k=0}^{\infty} h(x_k,u_k)\label{eq:Constraints1InfCost}\\
\text{s.t. }
   &x_{k+1}=f(x_k,u_k),~\forall k\geq 0 \label{eq:Constraints1Inf}\\
   &x_0=x_S,\label{eq:Constraints2Inf}\\
   &x_k \in \mathcal{X},~u_k \in \mathcal{U},~\forall k\geq 0\label{eq:Constraints3Inf}
\end{align}
\end{subequations}
where equations (\ref{eq:Constraints1Inf}) and (\ref{eq:Constraints2Inf}) represent the system dynamics and the initial condition, and (\ref{eq:Constraints3Inf}) are the state and input constraints.
We assume that the stage cost $h(\cdot,\cdot)$ in equation (\ref{eq:Constraints1InfCost}) is continuous and it satisfies
\begin{equation}\label{eq:RunningCost}
\begin{aligned}
h(x_F,0) = 0~\textrm{and}~ h(x_t^j,u_t^j) \succ 0 ~ \forall ~ x_t^j \in&~\rr^n \setminus \{x_F\},\\
& u_t^j \in \rr^m\setminus \{0\},
\end{aligned}
\end{equation}
where the final state $x_F$ is assumed to be a feasible equilibrium for the unforced system (\ref{eq:system})
\begin{equation}\label{eq:Def_x_f}
\begin{aligned}
 f(x_F,0)=x_F.\\
\end{aligned}
\end{equation}
Throughout the paper we assume that a local optimal solution to Problem (\ref{eq:ConstraintsInf}) exists and it is denoted as:
\begin{equation}\label{eq:Optimaltask}
\begin{aligned}
{\bf{x}}^{*} ~ = ~ [{{x}}^{*}_0,~{{x}}^{*}_1,~...,~{{x}}^{*}_t,~...],&&\\
{\bf{u}}^{*} ~ = ~ [{{u}}^{*}_0,~{{u}}^{*}_1,~...,~{{u}}^{*}_t,~...].&&\\
\end{aligned}
\end{equation}

\begin{remark}
By assumption, the stage cost $h(\cdot,\cdot)$ in (\ref{eq:RunningCost}) is continuous, strictly positive and zero at $x_F$. Thus, an optimal solution to (\ref{eq:ConstraintsInf}) converges to the final point $x_F$, i.e. $\lim_{t \to \infty} {{x}}^{*}_t =  x_F$. \\
\end{remark}

\begin{remark}
In practical applications each iteration has a finite-time duration.
It is common in the literature to adopt an infinite time formulation at each iteration for the sake of simplicity. We follow such an approach in this paper. Our choice does not affect the practicality of the proposed method.
\end{remark}

Next we introduce the definition  of the  sampled safe set and of the iteration cost. Both which will be used later to guarantee stability and feasibility of the learning MPC.

\subsection{Sampled Safe Set}

\begin{definition}[one-step controllable set to  the set $\mathcal{S}$]
For the system~(\ref{eq:system}) we denote the \emph{one-step controllable set to  the set $\mathcal{S}$} as
\end{definition}
\begin{equation}\label{eq:Pred}
\begin{aligned}
\mathcal{K}_1(\mathcal{S}) = \Pre(\mathcal{S}) \cap\mathcal{X}.
\end{aligned}
\end{equation}
where
\begin{equation}\label{eq:Pred0}
\begin{aligned}
\Pre(\mathcal{S})\triangleq \{x\in \rr^n~:~\exists u\in\mathcal{U}\text{ s.t. } f(x,u)\in \mathcal{S}\}.
\end{aligned}
\end{equation}
$\mathcal{K}_1(\mathcal{S})$ is the set of states which can be driven into the target set $\mathcal{S}$ in one time step while satisfying input and state constraints. $N$-step controllable sets are defined by iterating $\mathcal{K}_1(\mathcal{S})$  computations.

\begin{definition}[$N$-Step Controllable Set $\tKN(\mathcal{S})$] \index{$N$-Step Controllable Set}
\label{def:Ncset}
For a given target set $\mathcal{S}\subseteq \mathcal{X}$, the $N$-step controllable set $\tKN(\mathcal{S})$ of the system \eqref{eq:system} subject to the constraints~\eqref{eq:inv_constraints} is defined recursively as:
\begin{equation}
\label{eq:Nsteprecf}
\mathcal{K}_j(\mathcal{S}) \triangleq \Pre(\mathcal{K}_{j-1}(\mathcal{S}))\cap\mathcal{X},~~ \mathcal{K}_{0}(\mathcal{S})=\mathcal{S},~~~~ j\in\{1,\ldots,N\}
\end{equation}
\end{definition}
From Definition~\ref{def:Ncset}, all states $x_0$ of the system \eqref{eq:system} belonging to the $N$-Step Controllable Set $\tKN(\mathcal{S})$ can be driven, by a suitable control sequence, to the target set $\mathcal{S}$ in $N$ steps,
while satisfying input and state constraints.

\begin{definition}[Maximal Controllable Set $\tKinf(\mathcal{O})$] \label{def:maxContrSet}\index{Maximal Controllable Set}
For a given
target set $\mathcal{O}\subseteq \mathcal{X}$, the maximal controllable set
$\tKinf(\mathcal{O})$ for system \eqref{eq:system} subject to the constraints in \eqref{eq:inv_constraints} is the union of all $N$-step controllable sets $\tKN(\mathcal{O})$ contained in $\mathcal{X}$ ($N\in \mathbb{N}$).
\end{definition}

We will use controllable sets $\tKN(\mathcal{O})$ where the target $\mathcal{O}$ is a control invariant set~\cite{c16}.
They are special sets, since in addition to guaranteeing that from $\tKN(\mathcal{O})$ we reach $\mathcal{O}$ in $N$ steps,
one can ensure that once it has reached $\mathcal{O}$, the system can stay there at all future time instants. These sets are called stabilizable set.

\begin{definition}[$N$-step (Maximal) Stabilizable Set] \label{def:maxAttrSet}
For a given control invariant set $\mathcal{O}\subseteq \mathcal{X}$, the $N$-step (maximal)
stabilizable set of the system~\eqref{eq:system} subject to the constraints~\eqref{eq:inv_constraints}
is the $N$-step (maximal) controllable set $\tKN(\mathcal{O})$ ($\tKinf(\mathcal{O})$).
\end{definition}

Note that $x_F$ in (\ref{eq:Def_x_f}) is a control invariant since it is an equilibrium point. Therefore $\mathcal{K}_{\infty}(x_F)$ is the maximal stabilizable set to $x_F$.

Since the computation of Pre-set is numerically challenging for nonlinear systems, there is extensive literature on how to obtain an approximation (often conservative) of the maximal stabilizable set \cite{c26}.

In this paper we exploit the iterative nature of the control design. We notice that for every $k$-th iteration that successfully steers the system to the terminal point $x_F$ (i.e. $\forall k: \lim_{t \rightarrow \infty} x_t^k = x_F$), the state trajectory ${\bf{x}}^k$ is a subset of the maximal stabilizable set, i.e. $x_t^k \in \mathcal{K}_{\infty}(x_F), \forall t \geq 0$. Thus, we
define the \emph{sampled Safe Set} $\mathcal{SS}^j$ at iteration $j$ as
\begin{equation}\label{eq:SS}
\begin{aligned}
\mathcal{SS}^j = \textrm{}\left\{\bigcup_{i \in M^j} \bigcup_{t=0}^{\infty} x_t^i \right\}
\end{aligned}
\end{equation}
where
\begin{equation}\label{eq:M}
\begin{aligned}
M^j = \textrm{} \Big\{ k \in [0,j] : \lim_{t \to \infty} x_t^k = x_F \Big\}.
\end{aligned}
\end{equation}
$\mathcal{SS}^j$ is the collection of all state trajectories at iteration $i$ for $i\in M^j$. $M^j$ in equation (\ref{eq:M}) is the set of indexes $k$ associated with successful iterations $k$ for $k\leq j$.

From (\ref{eq:M}) we have that $M^i \subseteq M^j, \forall i \leq j$, which implies that
\begin{equation}\label{eq:SS_subset}
\begin{aligned}
\mathcal{SS}^i \subseteq \mathcal{SS}^j, \forall i \leq j.
\end{aligned}
\end{equation}

\begin{remark}
Note that $\mathcal{SS}^j$ can be interpreted as a sampled subset of the maximal stabilizable set $\tKinf(x_F)$ as for every point in the set, there exists a feasible control action which satisfies the state constraints and steers the state towards $x_F$.
\end{remark}

Lastly, we introduce the definition of successor set.
	\begin{definition}[one-step successor set from  the set $\mathcal{S}$]
		For the system~(\ref{eq:system}) we denote the \emph{one-step successor set from  the set $\mathcal{S}$} as
	\end{definition}
	\begin{equation}
	\begin{aligned}
	\mathcal{R}_1(\mathcal{S}) = \Succ(\mathcal{S}) \cap\mathcal{X}.
	\end{aligned}
	\end{equation}
	where
	\begin{equation}
	\begin{aligned}
	\Succ(\mathcal{S})\triangleq \{x\in \rr^n~:~\exists x(0) \in & ~\mathcal{S}, \exists u\in\mathcal{U} \\ & \text{ s.t. } f(x(0),u) = x \}.
	\end{aligned}
	\end{equation}

\subsection{Iteration Cost}
At time $t$ of the $j$-th iteration the cost-to-go associated with the closed loop trajectory (\ref{eq:sequenceX}) and input sequence (\ref{eq:sequenceU}) is defined as

\begin{equation}\label{eq:Functional}
\begin{aligned}
J_{t\rightarrow \infty}^j(x_t^j) = ~ \sum\limits_{k=t}^{\infty} h(x_k^j,u_k^j),
\end{aligned}
\end{equation}
where $h(\cdot,\cdot)$ is the stage cost of the problem (\ref{eq:ConstraintsInf}).
We define the \emph{$j$-th iteration cost}  as the cost (\ref{eq:Functional}) of the $j$-th trajectory at time $t=0$,
\begin{equation}\label{eq:Performance}
\begin{aligned}
J_{0\rightarrow \infty}^j(x_0^j) = ~ \sum\limits_{k=0}^{\infty} h(x_k^j,u_k^j).
\end{aligned}
\end{equation}
$J_{0\rightarrow \infty}^j(x_0^j)$ quantifies the controller performance at each $j$-th iteration.

\begin{remark}
	In equations (\ref{eq:Functional})-(\ref{eq:Performance}), $x_k^j$ and $u_k^j$ are the realized state and input at the $j$-th iteration, as defined in (\ref{eq:sequence}).
\end{remark}

\begin{remark}
At each $j$-th iteration the system is initialized at the same starting point $x_0^j = x_S$; thus we have $J_{0\rightarrow \infty}^j(x_0^j) = J_{0\rightarrow \infty}^j(x_S)$.
\end{remark}

Finally, we define the function $Q^j(\cdot)$, defined over the sample safe set $\mathcal{SS}^j$ as:
\begin{equation}\label{eq:Qfunction}
\begin{aligned}
Q^j(x) = \begin{cases} \min\limits_{ (i,t) \in F^j(x)} J^i_{t\rightarrow \infty}(x), & \mbox{if } x \in \mathcal{SS}^j \\
~~~+\infty, & \mbox{if } x \notin \mathcal{SS}^j \end{cases},
\end{aligned}
\end{equation}
where $F^j(\cdot)$ in (\ref{eq:Qfunction}) is defined as
\begin{equation}\label{eq:Qfunction1}
\begin{aligned}
F^j(x) = \textrm{} \Big\{ (i,t) : i \in [0,j],~t\geq 0 ~&\textrm{with}~ x_t^i = x; \\
&~\textrm{for}~x_t^i \in \mathcal{SS}^j \Big\}.
\end{aligned}
\end{equation}
\begin{remark}
The function $Q^j(\cdot)$ in (\ref{eq:Qfunction}) assigns to every point in the sampled safe set, $SS^j$, the minimum cost-to-go along the trajectories in $\mathcal{SS}^j$ i.e.,
\begin{equation}\label{eq:SStraj}
\begin{aligned}
\forall x \in SS^j, Q^j(x) = J^{i^*}_{t^*\rightarrow \infty}(x) = \sum\limits_{k=t^*}^{\infty} h(x_k^{i^*},u_k^{i^*}),
\end{aligned}
\end{equation}
where the indices pair $(i^*,t^*)$ is function of $x$ and it is the minimizer in (\ref{eq:Qfunction}):
\begin{equation}\label{eq:argMin}
\begin{aligned}
(i^*,t^*) =\mathop{\mathrm{argmin}}\limits_{ (i,t) \in F^j(x)} J^i_{t\rightarrow \infty}(x), ~~\mbox{for } x \in \mathcal{SS}^j.
\end{aligned}
\end{equation}
\end{remark}

In the next section we exploit the fact that at each iteration we solve the same problem to design a controller that guarantees a non-increasing iteration cost (i.e. $J_{0\rightarrow \infty}^j(\cdot) \leq J_{0\rightarrow \infty}^{j-1}(\cdot)$).

\addtolength{\textheight}{-3cm}   

\section{LMPC CONTROL DESIGN}
In this section we present the
design of the proposed Learning Model Predictive Control (LMPC). We first assume that there exists an iteration where the LMPC is feasible at all time instants.
Then, we prove that the proposed LMPC is guaranteed to be recursively feasible, i.e., feasible at all time instants of every successive iteration. Moreover, the trajectories from previous iterations are used to guarantee non-increasing iterations cost between two successive iterations.

\subsection{LMPC Formulation}

The LMPC tries to compute a solution to the infinite time optimal control problem (\ref{eq:ConstraintsInf}) by solving at time $t$ of iteration $j$ the finite time constrained optimal control problem

\begin{subequations}\label{eq:Constraints}
\begin{align}
&J_{t\rightarrow t+N}^{\scalebox{0.4}{LMPC},j}(x_t^j)=\min_{u_{t|t},\ldots,u_{t+N-1|t}} \bigg[  \sum_{k=t}^{t+N-1}  h(x_{k|t},u_{k|t}) +\notag\\
&~~~~~~~~~~~~~~~~~~~~~~~~~~~~~~~~~~~~~~~~~~~~~~+ Q^{j-1}(x_{t+N|t}) \bigg]\\
&\text{s.t. }\notag \\
   &~~~~x_{k+1|t}=f(x_{k|t},u_{k|t}),~\forall k \in [t, \ldots, t+N-1] \label{eq:Constraints1}\\
   &~~~~x_{k|t} \in \mathcal{X}, ~ u_{k|t} \in \mathcal{U},~\forall k \in [t, \ldots, t+N-1] \label{eq:Constraints4}\\
   &~~~~x_{t+N|t} \in ~\mathcal{SS}^{j-1},\label{eq:Constraints5} \\
   &~~~~x_{t|t}=x_t^j,\label{eq:Constraints2}
\end{align}
\end{subequations}
where (\ref{eq:Constraints1}) and (\ref{eq:Constraints2}) represent the system dynamics and initial condition, respectively. The state and input constraints are given by (\ref{eq:Constraints4}). Constraint (\ref{eq:Constraints5}) forces the terminal state into the set $\mathcal{SS}^{j-1}$ defined in equation (\ref{eq:SS}).\\
Let
\begin{equation}\label{eq:OptimalSolutionMPC}
\begin{aligned}
{\bf{u}}^{*,j}_{t:t+N|t}  &= [u_{t|t}^{*,j}, \ldots, u_{t+N-1|t}^{*,j}]\\
{\bf{x}}^{*,j}_{t:t+N|t} &= [x_{t|t}^{*,j}, \ldots, x_{t+N|t}^{*,j}]
\end{aligned}
\end{equation}
be the optimal solution of (\ref{eq:Constraints}) at time $t$ of the $j$-th iteration and $J_{t\rightarrow t+N}^{\scalebox{0.4}{LMPC},j}(x_t^j)$ the corresponding optimal cost. Then, at time $t$ of the iteration $j$, the first element of ${\bf{u}}^{*,j}_{t:t+N|t}$ is applied to the system (\ref{eq:system})
\begin{equation}\label{eq:MPC}
\begin{aligned}
u_t^j = u_{t|t}^{*,j}.
\end{aligned}
\end{equation}
The finite time optimal control problem (\ref{eq:Constraints}) is solved at time $t+1$, based on the new state $x_{t+1|t+1} = x_{t+1}^j$, yielding a \textit{moving} or \textit{receding horizon} control strategy.

\begin{assumption}
\label{ass1}
At iteration $j=1$ we assume that  $\mathcal{SS}^{j-1}= \mathcal{SS}^0$ is a non-empty set and that the trajectory ${\bf{x}}^0 \in \mathcal{SS}^0$ is feasible and convergent to $x_F$.
\end{assumption}

Assumption~\ref{ass1} is not restrictive in practice for a number of applications. For instance, with race cars one can always run a path following controller at very low speed to obtain a feasible state and input sequence.

In the next section we prove that, under Assumption~\ref{ass1}, the LMPC (\ref{eq:Constraints}) and (\ref{eq:MPC}) in closed loop with system (\ref{eq:system}) guarantees recursively feasibility and stability, and non-increase of the iteration cost at each iteration.

\begin{remark}
From (\ref{eq:SS}), $\mathcal{SS}^j$ at the $j$-th iteration is the set of all successful trajectories performed in the first $j$ trials. We assume that these trajectories can be recorded and stored at each iteration. Checking if a state is  in $\mathcal{SS}^j$ is  a simple search. However, the optimization problem~(\ref{eq:Constraints}) becomes challenging to solve even in the linear case due to the integer nature of the constraints~(\ref{eq:Constraints5}). In Section VI.A we comment on practical approaches to improve the computational  time to solve~(\ref{eq:Constraints}).
\end{remark}

\subsection{Recursive feasibility and stability}

As mentioned in Section II, for every point in the set $\mathcal{SS}^j$ there exists a control sequence that can drive the system to the terminal point $x_F$. 
The properties of $\mathcal{SS}^j$ and $Q^j(\cdot)$ are used in the next proof to show recursive feasibility and asymptotic stability of the equilibrium point $x_F$.\\

\begin{theorem}
Consider system (\ref{eq:system}) controlled by the \mbox{LMPC}
controller (\ref{eq:Constraints}) and (\ref{eq:MPC}).
Let $\mathcal{SS}^j$ be the sampled safe set at iteration $j$ as defined in (\ref{eq:SS}). Let assumption 1 hold, then the LMPC (\ref{eq:Constraints}) and (\ref{eq:MPC}) is feasible for all $t \geq 0$ and at every iteration $j\geq1$.
Moreover, the equilibrium point $x_F$ is asymptotically stable for the closed loop system (\ref{eq:system}), \eqref{eq:Constraints} and (\ref{eq:MPC}) at every iteration $j\geq1$.
\end{theorem}

\textbf{Proof:}
The proof follows from standard MPC arguments.\\
By assumption $\mathcal{SS}^0$ is non empty. From (\ref{eq:SS_subset}) we have that $\mathcal{SS}^{0} \subseteq \mathcal{SS}^{j-1} ~ \forall j \geq 1$, and consequently $\mathcal{SS}^{j-1}$ is a non empty set. In particular, there exists a feasible trajectory ${\bf{x}}^0 \in \mathcal{SS}^0 \subseteq \mathcal{SS}^{j-1}$. From (\ref{eq:Initial_state}) we know that $x_0^j = x_S~\forall j \ge 0$. At time $t=0$ of the $j$-th iteration the $N$ steps trajectory
\begin{equation}
\begin{gathered}\label{eq:SolutionFeasible0}
	{\bf{x}}^0_{0:N}=[x_0^0,~x_1^0,~...,x_{N}^0] \in \mathcal{SS}^{j-1},
\end{gathered}
\end{equation}
and the related input sequence,
\begin{equation}
\begin{gathered}\label{eq:SolutionFeasible}
	[u_0^0,~u_1^0,~...,u_{N-1}^0],
\end{gathered}
\end{equation}
satisfy input and state constrains (\ref{eq:Constraints1})-(\ref{eq:Constraints4})-(\ref{eq:Constraints2}). Therefore (\ref{eq:SolutionFeasible0})-(\ref{eq:SolutionFeasible}) is a feasible solution to the LMPC (\ref{eq:Constraints}) and (\ref{eq:MPC}) at $t=0$ of the $j$-th iteration.\\
Assume that at time $t$ of the $j$-th iteration the LMPC (\ref{eq:Constraints}) and (\ref{eq:MPC}) is feasible and let ${\bf{x}}^{*,j}_{t:t+N|t}$ and ${\bf{u}}^{*,j}_{t:t+N|t}$
be the optimal trajectory and input sequence, as defined in (\ref{eq:OptimalSolutionMPC}). From (\ref{eq:Constraints2}) and (\ref{eq:MPC}) the realized state and input at time $t$ of the $j$-th iteration are given by
\begin{equation}\label{eq:BC}
\begin{aligned}
x_t^j = x_{t|t}^{*,j},\\
u_t^j = u_{t|t}^{*,j}.
\end{aligned}
\end{equation}
The terminal constraint (\ref{eq:Constraints5}) enforces $x^{*,j}_{t+N|t} \in \mathcal{SS}^{j-1}$ and, from (\ref{eq:SStraj}),
\begin{equation}\label{eq:OptimalIndex}
\begin{aligned}
Q^{j-1}(x_{t+N|t}^{*,j}) = J^{i^*}_{t^*\rightarrow \infty}(x_{t+N|t}^{*,j}) = \sum\limits_{k=t^*}^{\infty} h(x_k^{i^*},u_k^{i^*}).
\end{aligned}
\end{equation}
Note that $x_{t^*+1}^{i^*}=f(x_{t^*}^{i^*},u_{t^*}^{i^*})$ and, by the definition of $Q^j(\cdot)$ and $F^j(\cdot)$ in  (\ref{eq:Qfunction})-(\ref{eq:Qfunction1}), $x_{t^*}^{i^*}=x_{t+N|t}^{*,j}$. Since the state update in (\ref{eq:system}) and (\ref{eq:Constraints1}) are assumed identical we have that
\begin{equation}\label{eq:NoMissmatch}
x_{t+1}^j = x_{t+1|t}^{*,j}.
\end{equation}
At time $t+1$ of the $j$-th iteration the input sequence
\begin{equation}\label{eq:Feasible_1}
[u_{t+1|t}^{*,j},~u_{t+2|t}^{*,j},~...,~u_{t+N-1|t}^{*,j},~u_{t^*}^{i^*}],
\end{equation}
and the related feasible state trajectory
\begin{equation}\label{eq:Feasible_2}
[x_{t+1|t}^{*,j},~x_{t+2|t}^{*,j},~...,~x_{t+N-1|t}^{*,j},~x_{t^*}^{i^*},~x_{t^*+1}^{i^*}]
\end{equation}
satisfy input and state constrains (\ref{eq:Constraints1})-(\ref{eq:Constraints2})-(\ref{eq:Constraints4}). Therefore, (\ref{eq:Feasible_1})-(\ref{eq:Feasible_2}) is a feasible solution for the LMPC (\ref{eq:Constraints}) and (\ref{eq:MPC}) at time $t+1$.\\
We showed that at the $j$-th iteration, $\forall j \geq 1$ , \emph{(i):} the LMPC is feasible at time $t=0$ and  \emph{(ii):} if the LMPC is feasible at time $t$, then the LMPC is feasible at time $t+1$. Thus, we conclude by induction that the LMPC in (\ref{eq:Constraints}) and (\ref{eq:MPC}) is feasible $\forall j \geq 1$ and $t \geq 0$.

Next we use the fact the Problem (\ref{eq:Constraints}) is time-invariant at each iteration $j$ and we replace $J_{t\rightarrow t+N}^{\scalebox{0.4}{LMPC},j}(\cdot)$ with $J_{0\rightarrow N}^{\scalebox{0.4}{LMPC},j}(\cdot)$. In order to show the asymptotic stability of $x_F$ we have to show that the optimal cost, $J_{0\rightarrow N}^{\scalebox{0.4}{LMPC},j}(\cdot)$, is a Lyapunov function for the equilibrium point $x_F$ (\ref{eq:Def_x_f}) of the closed loop system (\ref{eq:system}) and (\ref{eq:MPC}) \cite{c16}. Continuity of $J_{0\rightarrow N}^{\scalebox{0.4}{LMPC},j}(\cdot)$ can be shown as in \cite{c12}. From (\ref{eq:Constraints1InfCost}), $J_{0\rightarrow N}^{\scalebox{0.4}{LMPC},j}(x) \succ 0 ~ \forall ~ x \in \rr^n \setminus \{x_F\}$ and $J_{0\rightarrow N}^{\scalebox{0.4}{LMPC},j}(x_F)=0$. Thus, we need to show that  $J_{0\rightarrow N}^{\scalebox{0.4}{LMPC},j}(\cdot)$ is decreasing along the closed loop trajectory.\\
From (\ref{eq:NoMissmatch}) we have $x_{t+1|t}^{*,j}=x_{t+1}^j$, which implies that
\begin{equation}\label{eq:LyapProof_01}
\begin{aligned}
J_{0\rightarrow N}^{\scalebox{0.4}{LMPC},j}(x_{t+1|t}^*) = J_{0\rightarrow N}^{\scalebox{0.4}{LMPC},j}(x_{t+1}^j).
\end{aligned}
\end{equation}
Given the optimal input sequence and the related optimal trajectory in (\ref{eq:OptimalSolutionMPC}), the optimal cost is given by
\begin{equation}\label{eq:LyapProof}
\begin{aligned}
J_{0\rightarrow N}^{\scalebox{0.4}{LMPC},j}(x_t^j&)=\min_{u_{t|t},\ldots,u_{t+N-1|t}} \bigg[  \sum_{k=0}^{N-1}  h(x_{k|t},u_{k|t}) +\\
&~~~~~~~~~~~~~~~~~~~~~~~~~~~~~~~~~+ Q^{j-1}(x_{N|t}) \bigg] = \\
= h(x_{t|t}^{*,j}&,u_{t|t}^{*,j}) + \sum_{k=1}^{N-1}  h(x^{*,j}_{t+k|t},u^{*,j}_{t+k|t}) + Q^{j-1}(x^{*,j}_{t+N|t}) =\\
= h(x_{t|t}^{*,j}&,u_{t|t}^{*,j}) + \sum_{k=1}^{N-1}  h(x^{*,j}_{t+k|t},u^{*,j}_{t+k|t}) + J^{i^*}_{t^*\rightarrow \infty}(x^{*,j}_{t+N|t}) = \\
= h(x_{t|t}^{*,j}&,u_{t|t}^{*,j}) +  \sum_{k=1}^{N-1}  h(x^{*,j}_{t+k|t},u^{*,j}_{t+k|t}) + \sum_{k=t^*}^{\infty} h(x^{i^*}_{k},u^{i^*}_{k}) = \\
= h(x_{t|t}^{*,j}&,u_{t|t}^{*,j}) +  \sum_{k=1}^{N-1}  h(x^{*,j}_{t+k|t},u^{*,j}_{t+k|t}) + h(x^{i^*}_{t^*},u^{i^*}_{t^*}) ~ + \\
&~~~~~~~~~~~~~~~~~~~~~~~~~~~~~~~~~~~~~~+ Q^{j-1}(x^{i^*}_{t^*+1}) \geq \\
\geq& h(x_{t|t}^{*,j},u_{t|t}^{*,j}) + J_{0\rightarrow N}^{\scalebox{0.4}{LMPC},j}(x_{t+1|t}^{*,j}),
\end{aligned}
\end{equation}
where $(i^*, t^*)$ is defined in (\ref{eq:argMin}).

Finally, from equations (\ref{eq:MPC}), (\ref{eq:BC}) and (\ref{eq:LyapProof_01})-(\ref{eq:LyapProof}) we conclude that the optimal cost is a decreasing Lyapunov function along the closed loop trajectory,
\begin{equation}\label{eq:LyapProof2}
\begin{aligned}
J_{0\rightarrow N}^{\scalebox{0.4}{LMPC},j}(x_{t+1}^j)-J_{0\rightarrow N}^{\scalebox{0.4}{LMPC},j}(x_{t}^j) \leq - h&(x_{t}^{j},u_{t}^{j}) < 0, \\
\forall~x_t^j \in R^n& \setminus \{x_F\},~ \forall~u_t^j \in R^m \setminus \{0\} \\
\end{aligned}
\end{equation}
Equation (\ref{eq:LyapProof2}), the positive definitiveness of $h(\cdot)$ and the continuity of $J_{0\rightarrow N}^{\scalebox{0.4}{LMPC},j}(\cdot)$ imply that $x_F$ is asymptotically stable. \cvd

\subsection{Convergence properties}
In this section we assume that the LMPC (\ref{eq:Constraints}) and (\ref{eq:MPC}) converges to a steady state trajectory. We show two results. First, the $j$-th iteration cost $J_{0\rightarrow \infty}^{j}(\cdot)$ does not increase as $j$ increases. Second, the steady state trajectory is a local optimal solution to an approximation of the infinite horizon control problem (\ref{eq:ConstraintsInf}). We use the fact the Problem (\ref{eq:Constraints}) is time-invariant at each iteration $j$ and we replace $J_{t\rightarrow t+N}^{\scalebox{0.4}{LMPC},j}(\cdot)$ with $J_{0\rightarrow N}^{\scalebox{0.4}{LMPC},j}(\cdot)$.\\

\begin{theorem}
\label{th2}
Consider system (\ref{eq:system}) in closed loop with the LMPC controller (\ref{eq:Constraints}) and (\ref{eq:MPC}).
Let $\mathcal{SS}^j$ be the sampled safe set at the $j$-th iteration as defined in (\ref{eq:SS}). Let assumption 1 hold, then the iteration cost $J_{0\rightarrow \infty}^{j}(\cdot)$ does not increase with the iteration index $j$.
\end{theorem}

\textbf{Proof:}
First, we find a lower bound on the $j$-th iteration cost $J_{0\rightarrow \infty}^{j}(\cdot), ~\forall ~j>0$. Consider the realized state and input sequence (\ref{eq:sequence}) at the $j$-th iteration, which collects the first element of the optimal state and input sequence to the LMPC (\ref{eq:Constraints}) and (\ref{eq:MPC}) at time $t$, $~\forall t \geq 0$, as shown in (\ref{eq:BC}).
By the definition of the iteration cost in (\ref{eq:Functional}), we have
\begin{equation}\label{eq:ProofImprovingA}
\begin{aligned}
J_{0\rightarrow \infty}^{j-1}(x_S) &= \sum\limits_{t=0}^{\infty} h(x_t^{j-1},u_t^{j-1}) = \\
&=\sum\limits_{t=0}^{N-1} h(x_t^{j-1},u_t^{j-1}) + \sum\limits_{t=N}^{\infty} h(x_t^{j-1},u_t^{j-1}) \geq \\
&\geq \sum\limits_{t=0}^{N-1} h(x_t^{j-1},u_t^{j-1}) + Q^{j-1}(x_N^{j-1}) \geq \\
&\geq \min_{u_0,\ldots,u_{N-1}} \left[ \sum_{k=0}^{N-1}  h(x_k,u_k) + Q^{j-1}(x_N) \right] = \\
&= J_{0\rightarrow N}^{\scalebox{0.4}{LMPC},j}(x_0^{j}).
\end{aligned}
\end{equation}
Then we notice that, at the $j$-th iteration, the optimal cost of the LMPC (\ref{eq:Constraints}) and (\ref{eq:MPC}) at $t=0$, $J_{0\rightarrow N}^{\scalebox{0.4}{LMPC},j}(x_0^{j})$, can be upper bounded along the realized trajectory (\ref{eq:sequence}) using (\ref{eq:LyapProof2})
\begin{equation}\label{eq:ProofImprovingB}
\begin{aligned}
J_{0\rightarrow N}^{\scalebox{0.4}{LMPC},j}(x_0^{j})  &\geq  h(x_0^j,u_0^j) + J_{0\rightarrow N}^{\scalebox{0.4}{LMPC},j}(x_1^{j}) \phantom{\Big]} \geq\\
&\geq  h(x_0^j,u_0^j) + h(x_1^j,u_1^j)  + J_{0\rightarrow N}^{\scalebox{0.4}{LMPC},j}(x_2^{j}) \phantom{\Big]} \geq\\
&\geq\lim_{t \to \infty} \left[\sum_{k=0}^{t-1} h(x_k^j,u_k^j) + J_{0\rightarrow N}^{\scalebox{0.4}{LMPC},j}(x_t^{j}) \right]. \\
\end{aligned}
\end{equation}
From Theorem 1 $x_F$ is asymptotically stable for the closed loop system (\ref{eq:system}) and (\ref{eq:MPC}) (i.e. $\lim_{t \to \infty} x_t^{j} = x_F$), thus by continuity of $h(\cdot, \cdot)$
\begin{equation}\label{eq:LimProof}
	\begin{aligned}
	\lim_{t \rightarrow \infty} J_{0 \rightarrow N}^{\scalebox{0.5}{LMPC},j}(x_t) = J_{0 \rightarrow N}^{\scalebox{0.4}{LMPC},j}(x_F) = 0.
	\end{aligned}
\end{equation}
From equations (\ref{eq:ProofImprovingB})-(\ref{eq:LimProof})
\begin{equation}\label{eq:ProofImprovingC}
\begin{aligned}
J_{0\rightarrow N}^{\scalebox{0.4}{LMPC},j}(x_0^{j}) \geq \sum_{k=0}^{\infty} h(x_k^j,u_k^j) = J_{0\rightarrow \infty}^{j}(x_S),
\end{aligned}
\end{equation}
and finally from equations (\ref{eq:ProofImprovingA}) and (\ref{eq:ProofImprovingC}) we conclude that
\begin{equation}\label{eq:ProofImprovingF}
\begin{aligned}
J_{0\rightarrow \infty}^{j-1}(x_S) \geq J_{0\rightarrow N}^{\scalebox{0.4}{LMPC},j}(x_0^{j}) \geq  J_{0\rightarrow \infty}^{j}(x_S),
\end{aligned}
\end{equation}
thus the iteration cost is non-increasing. \cvd

Next,  we assume that the LMPC (\ref{eq:Constraints}) and (\ref{eq:MPC}) converges to a steady state trajectory $x_0^\infty,~x_1^\infty,\ldots$.
We try to answer the following question: ``What is the link between such steady state trajectory and an optimal solution to (\ref{eq:ConstraintsInf})?". We introduce the following finite time optimal control problem closely linked to Problem~\eqref{eq:ConstraintsInf},
\begin{subequations}\label{eq:FiniteHorizon}
	\begin{align}
	&\tilde{J}_{t\rightarrow t+T}^*(x_t)=\min_{u_{0},\ldots,u_{T-1}} \sum_{k=0}^{T-1}  h(x_{k},u_{k})+Q^{\infty}(x_T) \label{eq:FiniteHorizonCost}\\
	&\text{s.t. }\notag \\
	&~~~~x_{k+1}=f(x_{k},u_{k}),~\forall k \in [0, \ldots, T-1] \label{eq:FiniteHorizonDyn}\\
   &~~~~x_{k} \in \mathcal{X}, ~ u_{k} \in \mathcal{U},~\forall k \in [0, \ldots, T-1] \label{eq:FiniteHorizonConst}\\
	&~~~~x_{0}=x_t, ~x_{T} = x_{t+T}^{\infty},
	\end{align}
\end{subequations}
where the running cost in (\ref{eq:FiniteHorizonCost}), the dynamic constraint in (\ref{eq:FiniteHorizonDyn}), the state and input constraints in (\ref{eq:FiniteHorizonConst}) are the same as in (\ref{eq:ConstraintsInf}).

\begin{remark} \label{remark:Optimality}
Compare Problem (\ref{eq:FiniteHorizon}) with Problem (\ref{eq:Constraints}).
Problem (\ref{eq:FiniteHorizon}) uses an horizon $T$, possibly longer than the horizon $N$ of Problem (\ref{eq:Constraints}). Moreover, the terminal set of Problem (\ref{eq:FiniteHorizon}) is
a subset of the terminal set of Problem (\ref{eq:Constraints}).
Therefore, for $T=N$, every optimal solution to (\ref{eq:Constraints}) which is feasible Problem (\ref{eq:FiniteHorizon}) is also optimal.
\end{remark}

For the sake of simplicity we assume that Problem (\ref{eq:ConstraintsInf}) is strictly convex and
discuss the non-convex case in remark \ref{remark:nonconvex}.
\begin{assumption}
Problem (\ref{eq:ConstraintsInf}) is strictly convex.\\
\end{assumption}

\begin{theorem}
\label{th3}
Consider system (\ref{eq:system}) in closed loop with the LMPC controller (\ref{eq:Constraints}) and (\ref{eq:MPC}) with $N > 1$.
Let $\mathcal{SS}^j$ be the sampled safe set at the $j$-th iteration as defined in (\ref{eq:SS}). Let Assumptions 1-2 hold and assume that the LMPC controller (\ref{eq:Constraints}) and (\ref{eq:MPC}) converges to the steady state input ${\bf{u}}^\infty = \lim_{j \to \infty} {\bf{u}}^j$ and the steady state trajectory ${\bf{x}}^\infty = \lim_{j \to \infty} {\bf{x}}^j$, for iteration $j\rightarrow \infty$.
If $x_k^{\infty} \in \Int ( \Pre (x_{k+1}^{\infty}))$ and $x_{k+1}^{\infty} \in \Int( \Succ(x_{k}^{\infty}))$ for all $k \geq 0$, then $({\bf{x}}^\infty_{t:t+T}, {\bf{u}}^\infty_{t:t+T})$ is the optimizer of the finite horizon optimal control problem~(\ref{eq:FiniteHorizon}) with initial condition $x_t=x_t^\infty$ for all $t\geq0$ and for all $T>0$.
\end{theorem}

\textbf{Proof:}
By assumption,
system (\ref{eq:system}) in closed loop with the LMPC controller (\ref{eq:Constraints}) and (\ref{eq:MPC})
converges to a steady state trajectory ${\bf{x}^{\infty}}$. This implies that both the sampled safe set $SS^j $ and the terminal cost $Q^{j}(\cdot)$ converge at steady state, i.e,
for $j \rightarrow \infty$,  $\bf{x}^j \rightarrow {\bf{x}^{\infty}}$,  $SS^j \rightarrow SS^{\infty}$ and $Q^{j}(\cdot) \rightarrow Q^{\infty}(\cdot)$. From (\ref{eq:LyapProof2}) we have that
\begin{equation}\label{eq:CostBound}
\begin{aligned}
J_{0\rightarrow N}^{\scalebox{0.4}{LMPC},\infty}(x_{t}^{\infty}) &\geq h(x_{t}^{\infty},u_{t}^{\infty}) + J_{0\rightarrow N}^{\scalebox{0.4}{LMPC},\infty}(x_{t+1}^{\infty}) \phantom{\Big]} \geq \\
 \phantom{\Big]}\geq &h(x_{t}^{\infty},u_{t}^{\infty}) + h(x_{t+1}^{\infty},u_{t+1}^{\infty}) + J_{0\rightarrow N}^{\scalebox{0.4}{LMPC},\infty}(x_{t+2}^{\infty}) \geq\\
\geq & \left[ \sum_{k=0}^{T-1} h(x_{t+k}^{\infty},u_{t+k}^{\infty}) + J_{0\rightarrow N}^{\scalebox{0.4}{LMPC},\infty}(x_{t+T}^{\infty}) \right]~\forall~T>0. \\
\end{aligned}
\end{equation}
Since the terminal cost converges at steady state, we have that
\begin{equation}
\label{eq:CostBound02}
J_{0\rightarrow N}^{\scalebox{0.4}{LMPC},\infty}(x_{t+T}^{\infty})=
Q^{\infty}(x_{t+T}^{\infty}).
\end{equation}
From definition (\ref{eq:SS}), we have that ${\bf{x}^{\infty}} \in \mathcal{SS}^{\infty}$.
In equation~(\ref{eq:CostBound}), pick $T=N$ and from~(\ref{eq:CostBound02}) we have:
\begin{equation}\label{eq:CostBound1}
\begin{aligned}
J_{0\rightarrow N}^{\scalebox{0.4}{LMPC},\infty}(x_{t}^{\infty})
&\geq \sum_{k=0}^{N-1} h(x_{t+k}^{\infty},u_{t+k}^{\infty}) + Q^{\infty}(x_{t+N}^{\infty}).
\end{aligned}
\end{equation}
From ~(\ref{eq:CostBound1}) we conclude that the cost associated with the feasible state and input trajectory
\begin{equation}
\begin{aligned}\label{eq:Hamiltonian0}
	&{\bf{x}}^{\infty}_{t:t+N} = [x_{t}^{\infty},~x_{t+1}^{\infty},~...,~x_{t+N}^{\infty}] \\
	&{\bf{u}}^{\infty}_{t:t+N} = [u_{t}^{\infty},~u_{t+1}^{\infty},~...,~u_{t+N-1}^{\infty}]
\end{aligned}
\end{equation}
is a lower bound of the optimal cost $J_{0\rightarrow N}^{\scalebox{0.4}{LMPC},\infty}(x_{t}^{\infty})$. Therefore, (${\bf{x}}^{\infty}_{t:t+N}$, ${\bf{u}}^{\infty}_{t:t+N}$)
is an optimal solution to the LMPC (\ref{eq:Constraints})-\eqref{eq:MPC} for any $t$ and for $j \rightarrow \infty$.

From remark \ref{remark:Optimality} and from the above results, we have that $({\bf{x}}_{t:t+N}^{\infty}, {\bf{u}}_{t:t+N}^{\infty})$ is an optimal solution to the optimal control problem defined in (\ref{eq:FiniteHorizon}) with initial condition $x_t=x_t^{\infty}$ and $T=N$. The corresponding optimal cost is $\tilde J_{t\rightarrow t+N}^*(x_t^{\infty})$.

Next, we prove that ${\bf{x}}^{\infty}_{t:t+N+1}$ and ${\bf{u}}^{\infty}_{t:t+N+1}$ is the optimal solution to the finite time optimal control problem (\ref{eq:FiniteHorizon}) with initial condition $x_t=x_t^{\infty}$ and $T=N+1$. The corresponding optimal cost is $\tilde J_{t\rightarrow t+N+1}^*(x_t^{\infty})$. From time-invariance we focus on the case $t=0$ and refer to
Problem (\ref{eq:FiniteHorizon}) with initial condition $x_0=x_0^{\infty}$ and $T=N+1$ as
$J_{0\rightarrow N+1}^*(x_0^{\infty})$.\\
We proceed by contradiction and assume that the optimal solution to
problem $J_{0\rightarrow N+1}^*(x_0^{\infty})$ is
(${\tilde{x}}^{\infty}_{0:N+1}$, ${\tilde{u}}^{\infty}_{0:N+1}$) different from
(${\bf{x}}^{\infty}_{0:N+1},{\bf{u}}^{\infty}_{0:N+1}$).

Define $N$ feasible trajectories, for  $\alpha \in (0,1)$,
\begin{equation} \label{eq:NewTraj}
	\hat{x}_{0:N+1}^{i, \infty} = [x_0^{\infty}, \ldots, x_{i-1}^{\infty}, \alpha \tilde{x}^{\infty}_i +  (1 - \alpha) x_i^{\infty}, x_{i+1}^{\infty}, \ldots, x_{N+1}^{\infty}]
\end{equation}
 with $i = [1, \ldots, N]$.
By assumption $x_k^{\infty} \in \Int( \Pre(x_{k+1}^{\infty}))$ and $x_{k+1}^{\infty} \in \Int( \Succ(x_{k}^{\infty}))$ for all $k \geq 0$.
This implies that there exists an $\alpha >0$ such that the trajectory $\hat{x}_{0:N+1}^{i, \infty}$  and its related input sequence $\hat{u}_{0:N+1}^{i, \infty}$ are feasible for problem $J_{0\rightarrow N+1}^*(x_0^{\infty})$.

For easier readability we introduce the function $J_{0 \rightarrow N+1}(\cdot)$ which evaluates the cost of the $N+1$-steps trajectory:
\begin{equation}
	J_{0 \rightarrow N+1}({\bf{x}}_{0:N+1}^{\infty}) = \sum_{k =0}^{N} h(x_k^{\infty}, u_k^{\infty}) + Q(x_{N+1}^{\infty}).
\end{equation}
We notice that, by optimality of $({\bf{x}}^{\infty}_{t:t+N}, {\bf{u}}^{\infty}_{t:t+N})$ for all $t \geq 0$, we have \begin{subequations}
\begin{align}
J_{0 \rightarrow N}(\hat x^{i, \infty}_{0:N}) &>  J_{0 \rightarrow N} ({\bf{x}}_{0:N}^{\infty}),~~ \forall i \in [1, \ldots, N-1], \label{eq:CostTilda1} \\
J_{1 \rightarrow N+1}(\hat x^{i, \infty}_{1:N+1}) &>  J_{1 \rightarrow N+1} ({\bf{x}}_{1:N+1}^{\infty}),~~ \forall i \in [2, \ldots, N] \label{eq:CostTilda2}.
\end{align}
\end{subequations}
From (\ref{eq:CostTilda1}) we have
\begin{equation}\label{eq:Cost1proof}
\begin{aligned}
\sum_{k =0}^{N-1} h(\hat x_k^{i, \infty}, \hat u_k^{i, \infty}) + Q(\hat x_{N}^{i, \infty}) >  \sum_{k =0}^{N-1} & h(x_k^{\infty}, u_k^{\infty}) + Q(x_{N}^{\infty}) \\
&~~~\forall i \in [1, \ldots, N-1]. 
\end{aligned}
\end{equation}
Moreover, we know that $\hat x_N^{i, \infty} = x_N^{\infty}$ and $\hat u_N^{i, \infty} = u_N^{\infty}~ \forall i \in [1, \ldots, N-1]$, therefore by definition of the terminal cost \eqref{eq:Qfunction} and from \eqref{eq:Cost1proof} we have
\begin{equation}
\begin{aligned}
\sum_{k =0}^{N} h(\hat x_k^{i, \infty}, \hat u_k^{i, \infty}) + Q(\hat x_{N+1}^{i, \infty}) >  \sum_{k =0}^{N} &h(x_k^{\infty}, u_k^{\infty}) + Q(x_{N+1}^{\infty})
\\
&~~~\forall i \in [1, \ldots, N-1],
\end{aligned}
\end{equation}
which implies 
\begin{equation}\label{eq:CostStep1}
J_{0 \rightarrow N+1}(\hat x^{i, \infty}_{0:N+1}) >  J_{0 \rightarrow N+1} ({\bf{x}}_{0:N+1}^{\infty}),~~ \forall i \in [1, \ldots, N-1].
\end{equation}
Moreover, from the fact that $\hat x_0^{i, \infty} = x_0^{\infty}$ and $\hat u_0^{i, \infty} = u_0^{\infty},~ \forall i = [2, \ldots, N]$ and from (\ref{eq:CostTilda2}) we have
 \begin{equation}
 \begin{aligned}\label{eq:CostStep2}
 h(\hat x_0^{i, \infty}, \hat u_0^{i, \infty}) +& J_{1 \rightarrow N+1}(\hat x^{i, \infty}_{1:N+1}) > \\ &>h(x_0^{\infty}, u_0^{\infty}) + J_{1 \rightarrow N+1} ({\bf{x}}_{1:N+1}^{\infty}),\\&~~~~~~~~~~~~~~~~~~~~~~~~~~~~~~~~ \forall i \in [2, \ldots, N].
 \end{aligned}
\end{equation}
From (\ref{eq:CostStep1}) and (\ref{eq:CostStep2}) we conclude that
\begin{equation}
	 J_{0 \rightarrow N+1}(\hat x^{i, \infty}_{0:N+1}) >  J_{0 \rightarrow N+1} ({\bf{x}}_{0:N+1}^{\infty}),~~ \forall i \in [1, \ldots, N].
\end{equation}

Define the trajectory $\bar x_{0:N+1}^{\infty}$ as convex combination of ${\bf{x}}^{\infty}_{0:N+1}$ and the trajectories in \eqref{eq:NewTraj},
\begin{equation}\label{eq:TrajectoryConvexComb1}
\bar x_{0:N+1}^{\infty} = \sum_{i=1}^{N} \frac{1}{N} \Big(\frac{1}{2} \hat{x}^{i, \infty}_{0:N+1} + \frac{1}{2} {\bf{x}}^{\infty}_{0:N+1}\Big).
\end{equation}
From (\ref{eq:NewTraj}) we have that $\bar x_{0:N+1}^{\infty}$ can be expressed also as a convex combination of the optimal trajectory $\tilde{x}^{\infty}_{0:N+1}$ and ${\bf{x}}^{\infty}_{0:N+1}$,
\begin{equation}\label{eq:TrajectoryConvexComb2}
	\bar x_{0:N+1}^{\infty} = \frac{\alpha}{2N} \tilde{x}^{\infty}_{0:N+1} + \frac{2N-\alpha}{2N} {\bf{x}}^{\infty}_{0:N+1}.
\end{equation}

Concluding from (\ref{eq:TrajectoryConvexComb1}) and Assumption 2, we have that
\begin{equation} \label{eq:CostConvexComb1}
\begin{aligned}
J_{0 \rightarrow N+1} ({\bf{x}}_{0:N+1}^{\infty}) <  J_{0 \rightarrow N+1} (\bar x_{0:N+1}^{\infty}) < J_{0 \rightarrow N+1}  (\hat x_{0:N+1}^{k, \infty})
\end{aligned}
\end{equation}
where $k = \arg\max_{i \in [1, \ldots, N]}  J_{0 \rightarrow N+1}  (\hat x_{0:N+1}^{i, \infty})$. \\ Furthermore, from (\ref{eq:TrajectoryConvexComb2})  and Assumption 2, we have that
\begin{equation} \label{eq:CostConvexComb2}
	 J_{0 \rightarrow N+1}  (\tilde x_0^{\infty}) < J_{0 \rightarrow N+1} (\bar x_0^{\infty}) < J_{0 \rightarrow N+1} ({\bf{x}}_{0:N+1}^{\infty}).
\end{equation}
Finally, from (\ref{eq:CostConvexComb1}) and (\ref{eq:CostConvexComb2}) we have a contradiction and we conclude that (${\bf{x}}^{\infty}_{0:N+1}$, ${\bf{u}}^{\infty}_{0:N+1}$) is the optimal solution of the finite time optimal control problem~(\ref{eq:FiniteHorizon}) with initial condition $x_t=x_t^{\infty}$ and $T=N+1$.
The above procedure can be iterated for $T=N+2$,~$T=N+3$,\ldots which proves the Theorem.
\cvd

\begin{remark} \label{remark:nonconvex}
When problem \eqref{eq:ConstraintsInf} is non-convex, only local properties can be shown. In particular if one assumes that all local optimal solutions of \eqref{eq:ConstraintsInf} and \eqref{eq:Constraints}  are strict, then the proof of \textit{Theorem 3} could be modified to show local optimality of $({\bf{x}}^\infty_{0:T}, {\bf{u}}^\infty_{0:T})$ for the finite horizon optimal control problem $\tilde{J}_{0\rightarrow T}^*(x_S)$  for all $T>0$.
\end{remark}

\section{Examples}
\subsection{Constrained LQR controller}
In this section, we test the proposed LMPC on the following infinite horizon linear quadratic regulator with constraints (CLQR)
\begin{subequations}\label{eq:CLQR}
\begin{align}
J_{0\rightarrow \infty}^*(x_S)&=\min_{u_0, u_1,\ldots} \sum\limits_{k=0}^{\infty} \Big[ ||x_k||_2^2 + ||u_k||_2^2 \Big] \\
\textrm{s.t. }
   &x_{k+1}= \begin{bmatrix} 1 & 1 \\ 0 & 1 \end{bmatrix} x_k +  \begin{bmatrix} 0 \\ 1 \end{bmatrix} u_k,~\forall k\geq 0 \label{eq:CLQR1}\\
   &x_0=[-3.95 ~ ~ -0.05]^T,\label{eq:CLQR2}\\
   & \begin{bmatrix} -4 \\ -4 \end{bmatrix} \leq x_k \leq \begin{bmatrix} 4 \\ 4 \end{bmatrix} ~ \forall k\geq 0\label{eq:CLQR3} \\
   &-1 \leq u_k \leq 1 ~~\forall k\geq 0. \label{eq:CLQR4}
\end{align}
\end{subequations}

Firstly, we compute a feasible solution to (\ref{eq:CLQR}) using an open loop controller that drives the system close to the origin and, afterwards, an unconstrained LQR feedback controller. This feasible trajectory is used to construct the sampled safe set, $\mathcal{SS}^{0}$, and the terminal cost, $Q^{0}(\cdot)$, needed to initialize the first iteration of the LMPC (\ref{eq:Constraints}) and (\ref{eq:MPC}).

The LMPC (\ref{eq:Constraints}) and (\ref{eq:MPC}) is implemented with the quadratic running cost $h(x_k,u_k) = ||x_k||_2^2 + ||u_k||_2^2$, an horizon length $N = 4$, and the states and input constraints (\ref{eq:CLQR3})-(\ref{eq:CLQR4}). The LMPC (\ref{eq:Constraints}) and (\ref{eq:MPC}) is reformulated as a Mixed Integer Quadratic Programming and it is implemented in YALMIP \cite{c28} using the solver bonmin \cite{c32}. Each $j$-th iteration has an unknown fixed-time duration, $\tilde{t}_j$, defined as
\begin{equation}\label{eq:Terminate}
\begin{aligned}
\tilde{t}_j = \min \textrm{} \Big\{ t\in\zz_{0+}  : J_{0\rightarrow N}^{\scalebox{0.4}{LMPC},j}(x_t^j) \leq \epsilon \Big\}.
\end{aligned}
\end{equation}
with $\epsilon = 10^{-8}$.
Furthermore, each $j$-th closed loop trajectory is used to enlarge the sampled safe set used at the $j\text{+}1$-th iteration.

After $9$ iterations, the LMPC converges to steady state solution ${\bf{x}^{\infty}}={\bf{x}^{9}}$ with a tollerance of $\gamma$:
\begin{equation}\label{eq:Steady}
\begin{aligned}
\max_{t \in [0, \tilde{t}_9]} ||{{x}}^9_t-{{x}}^8_t||_2 < \gamma
\end{aligned}
\end{equation}
with $\gamma = 10^{-10}$. Table \ref{table:SS_Dimension} reports the number of points in the sampled safe set at each $j$-th iteration, until convergence is reached.

\textbf{\begin{table}[h!]
		\caption{Number of points in the sampled safe set.}\label{table:SS_Dimension}
		\centering
		\begin{tabular}{lll}
			\multicolumn{3}{l}{~~\textbf{Iteration}}{~~\textbf{Iteration Cost}}                   \\ \hline
			& $j = 1$    & ~~~~$62$ \\
			& $j =2$     & ~~~~$77$  \\
			& $j =3$     & ~~~~$92$  \\
			& $j =4$     & ~~~~$107$  \\
			& $j =5$     & ~~~~$122$  \\
			& $j =6$     & ~~~~$137$  \\
			& $j =7$     & ~~~~$152$  \\
			& $j =8$     & ~~~~$167$  \\
			& $j =9$     & ~~~~$182$  \\
		\end{tabular}
	\end{table}}
We observe that the iteration cost is non-increasing over the iterations and the LMPC (\ref{eq:Constraints}) and (\ref{eq:MPC}) improves the closed loop performance, as shown in Table \ref{table:Cost}.
\textbf{\begin{table}[h!]
\caption{Cost of the LMCPC at each $j$-th iteration}\label{table:Cost}
\centering
\begin{tabular}{lll}
\multicolumn{3}{l}{~~\textbf{Iteration}}{~~\textbf{Iteration Cost}}                   \\ \hline
 & $j = 0$    & ~~~~$57.1959612323$ \\
 & $j =1$     & ~~~~$49.9313760802$  \\
 & $j =2$     & ~~~~$49.9166093038$  \\
 & $j =3$     & ~~~~$49.9163668249$  \\
 & $j =4$     & ~~~~$49.9163602456$  \\
 & $j =5$     & ~~~~$49.9163600500$  \\
 & $j =6$     & ~~~~$49.9163600443$  \\
 & $j =7$     & ~~~~$49.9163600441$  \\
 & $j =8$     & ~~~~$49.9163600440$  \\
 & $j =9$     & ~~~~$49.9163600440$  \\
\end{tabular}
\end{table}}

We compare this steady state trajectory with the exact solution of the CLQR (\ref{eq:CLQR}), which is computed using the algorithm in \cite{c16}.
We analyze the deviation error,
\begin{equation}
\begin{gathered}\label{eq:deviation}
	\sigma_t = ||x_t^{\infty}-x^*_t||_2,
\end{gathered}
\end{equation}
which quantifies, at each time step $t$, the distance between the optimal trajectory ${\bf{x}^{*}}$ of the CLQR (\ref{eq:CLQR}) and steady state trajectory ${\bf{x}^{\infty}}$ at which the LMPC (\ref{eq:Constraints}) and (\ref{eq:MPC}) has converged. We notice that the maximum deviation error is $\max[\sigma_0, \dots, \sigma_{\tilde{t}_{\infty}}] = 1.62 \times 10^{-5}$, and that the $2$-norm of the difference between the exact optimal cost and the cost associated with the steady state trajectory is $||J_{0\rightarrow \infty}^*({{x}}^*_0)-J_{0\rightarrow \infty}^{\infty}(x_0^{\infty})||_2 = 1.565 \times 10^{-20}$. Therefore, we confirm that the LMPC (\ref{eq:Constraints}) and (\ref{eq:MPC}) has converged to a locally optimal solution that in the specific case is the global optimal solution which saturates both state and input constraints as the exact solution to the CLQR (\ref{eq:CLQR}).

\subsection{Dubins Car with Obstacle and Acceleration Saturation}
In this section, we test the proposed LMPC on the minimum time Dubins car problem \cite{c30} in discrete time. In this example, we add a known saturation limit on the acceleration in order to simulate the behavior of the friction circle \cite{gao2010predictive}, \cite{rajamani2011vehicle}. We control the car acceleration and steering. The controller's goal is to steer the system from the starting point $x_S$ to the unforced equilibrium point $x_F$. The minimum time optimal control problem is formulated as the following infinite time optimal control problem
\begin{subequations}\label{eq:Dubins}
	\begin{align}
	&J_{0\rightarrow \infty}^*(x_S)=\min_{\begin{smallmatrix} \theta_0, \theta_1,\ldots\\ a_0,a_1,\ldots \end{smallmatrix}} \sum\limits_{k=0}^{\infty} \mathds{1}_k  \label{eq:Dubins0}\\
	& {\textrm{s.t. }} \notag\\
	&x_k = \begin{bmatrix} z_{k+1} \\ y_{k+1} \\ v_{k+1} \end{bmatrix}= \begin{bmatrix} z_{k} \\ y_{k} \\ v_{k} \end{bmatrix} + \begin{bmatrix} v_k cos(\theta_k)\\ v_k sin(\theta_k)\\ a_k \end{bmatrix},~\forall k\geq 0 \label{eq:Dubins1}\\
	&x_0=x_S = [0~ ~ 0 ~~ 0]^T,\label{eq:Dubins2}\\
	&-s \leq a_k \leq s, ~~\forall k\geq 0 \label{eq:Dubins3}\\
	& \frac{(z_k - z_{obs})^2}{a_e^2} + \frac{(y_k - y_{obs})^2}{b_e^2} \geq 1, ~~ \forall k\geq 0.\label{eq:Dubins5}
	\end{align}
\end{subequations}
where the indicator function in (\ref{eq:Dubins0}) is defined as
\begin{equation}\label{eq:Indicator}
\begin{aligned}
\mathds{1}_k = \begin{cases} 1, & \mbox{if } x_k \neq {x}_F\\
0, & \mbox{if } x_k = {x}_F \end{cases}.
\end{aligned}
\end{equation}
In Equation (\ref{eq:Dubins3}), $s=1$ is the known acceleration saturation limit. Equations (\ref{eq:Dubins1})-(\ref{eq:Dubins2}) represent the dynamic constraint and the initial conditions, respectively. The state vector $x_k = [z_k, y_k, v_k]$ collects the car's position on the $Z-Y$ plane and the velocity, respectively. The inputs are the heading angle, $\theta_k$, and the acceleration command, $a_k$. Finally, (\ref{eq:Dubins5}) represents the obstacle constraint, enforcing the system trajectory to lie outside the ellipse centered at ($z_{obs}$,$y_{obs}$).

In order to find a local optimal solution to Problem (\ref{eq:Dubins}), we implemented the LMPC (\ref{eq:Constraints}) and (\ref{eq:MPC}) with the running cost $h(x_k, u_k) = \mathds{1}_k $ and constraints (\ref{eq:Dubins1})-(\ref{eq:Dubins5}).
We set $x_F = [54, 0, 0]^T$, $a_e=8$ and $b_e=6$.
At the $0$-th iteration, we computed a feasible trajectory that steers system (\ref{eq:Dubins}) from $x_0$ to $x_F$ using a brute force algorithm.
For efficient techniques to compute collision-free trajectories in the presence of obstacle we refer to \cite{bullo,frazzoli} and \cite{karaman2010incremental}. The feasible trajectory is used to construct the sampled safe set $\mathcal{SS}^{0}$, and the terminal cost, $Q^{0}(\cdot)$, needed to initialize the first iteration of the LMPC  (\ref{eq:Constraints}) and (\ref{eq:MPC}).

For this example, Problem~(\ref{eq:Constraints}) can be reformulated as a Mixed-Integer Quadratic Program (MIQP). Further details on its solution can be found in  Section VII.A.2.

\begin{figure}[h!]
	\centering
	\includegraphics[width=0.9\columnwidth]{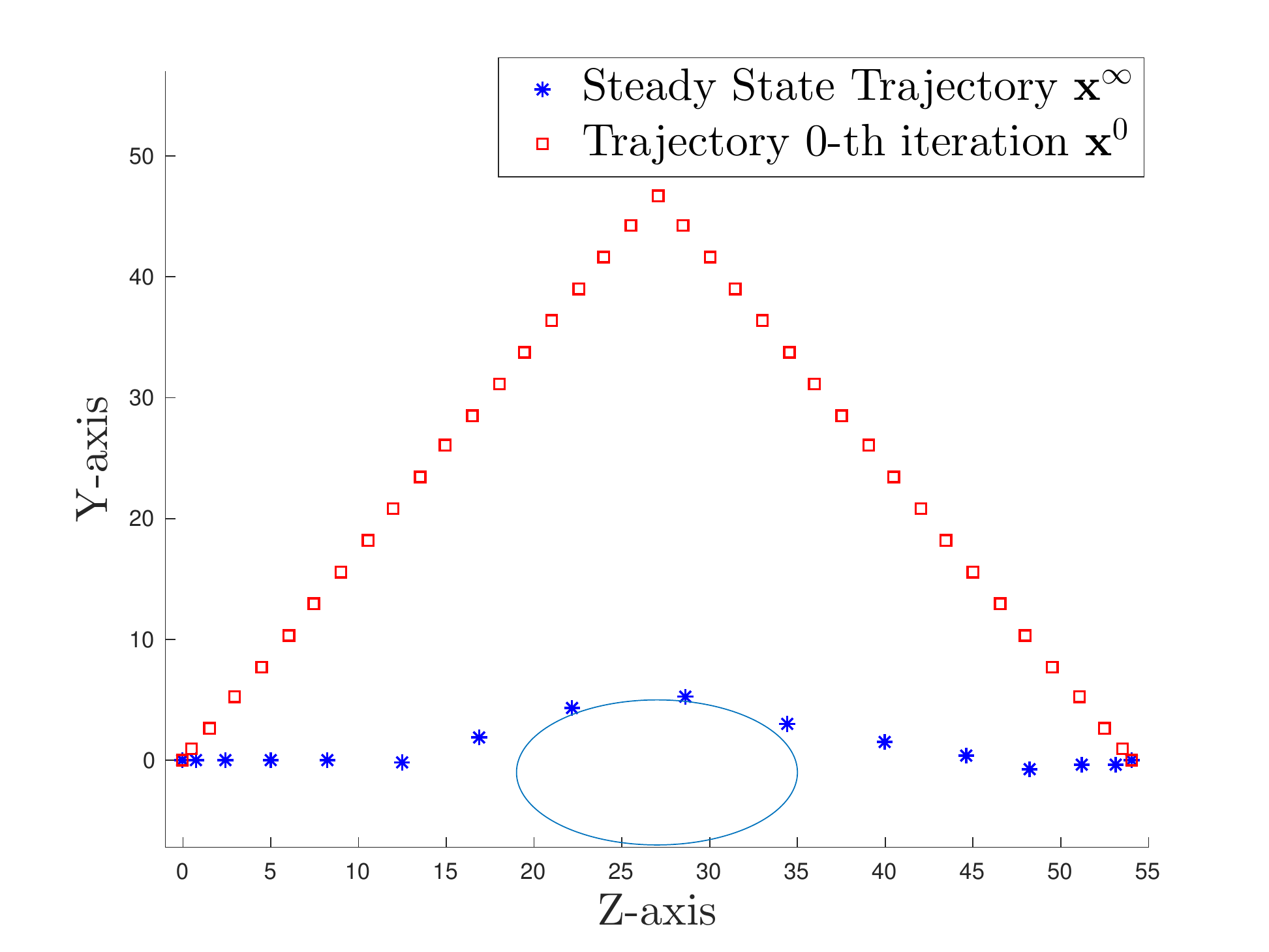}
	\caption{Comparison between the first feasible trajectory ${\bf{x}^0}$ and the steady state trajectory ${\bf{x}^{\infty}}$.}
	\label{Res2:DubinsSteadyState}
\end{figure}

\begin{figure}[h!]
	\centering
	\includegraphics[width=0.9\columnwidth]{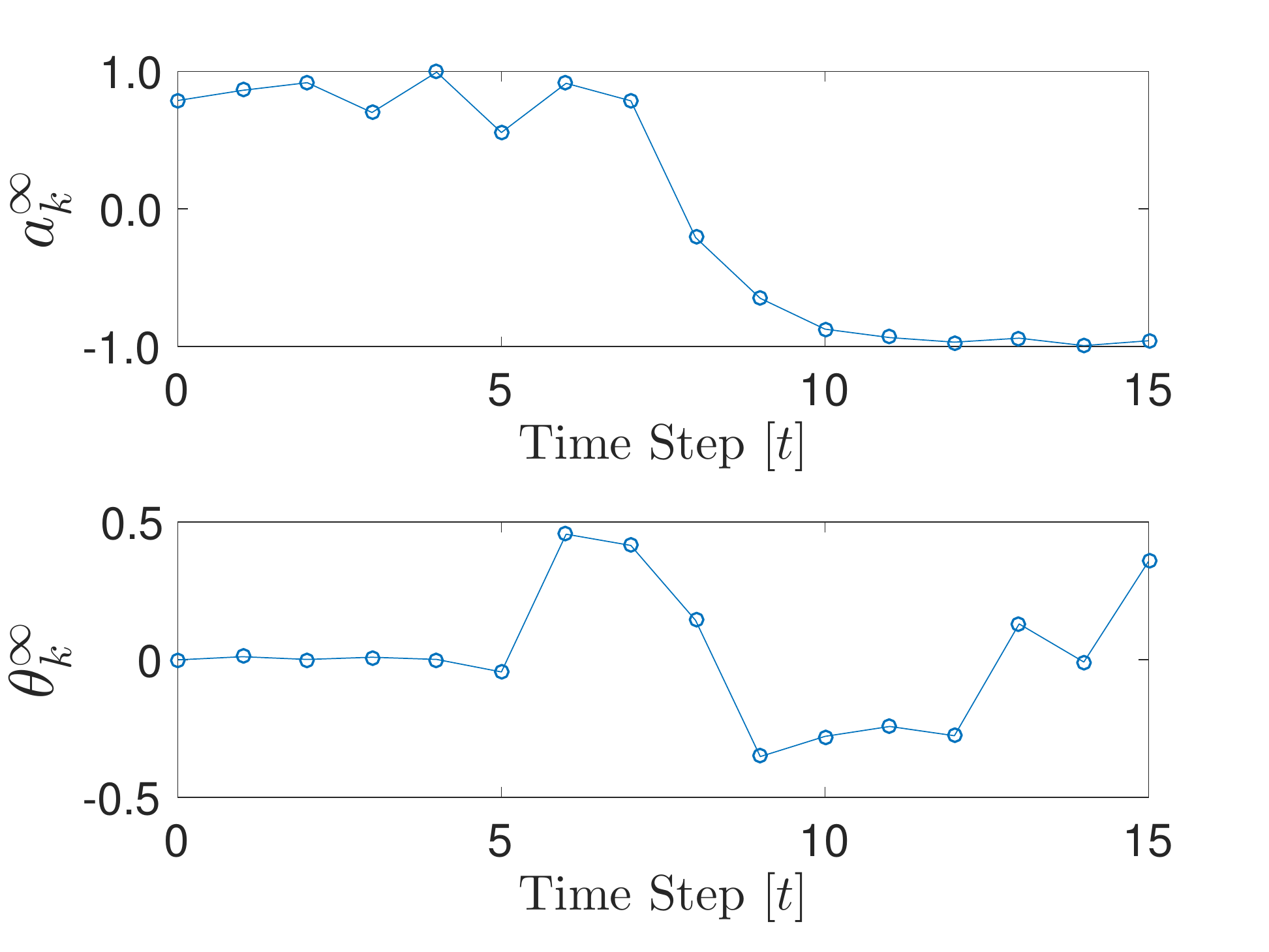}
	\caption{The acceleration $a_k^{\infty}$ and steering $\theta_k^{\infty}$ inputs associated with the steady state trajectory ${\bf{x}^{\infty}}$.}
	\label{Res2:DubinsSteadyStateInputs}
\end{figure}

\begin{figure}[h!]
	\centering
	\includegraphics[width=0.9\columnwidth]{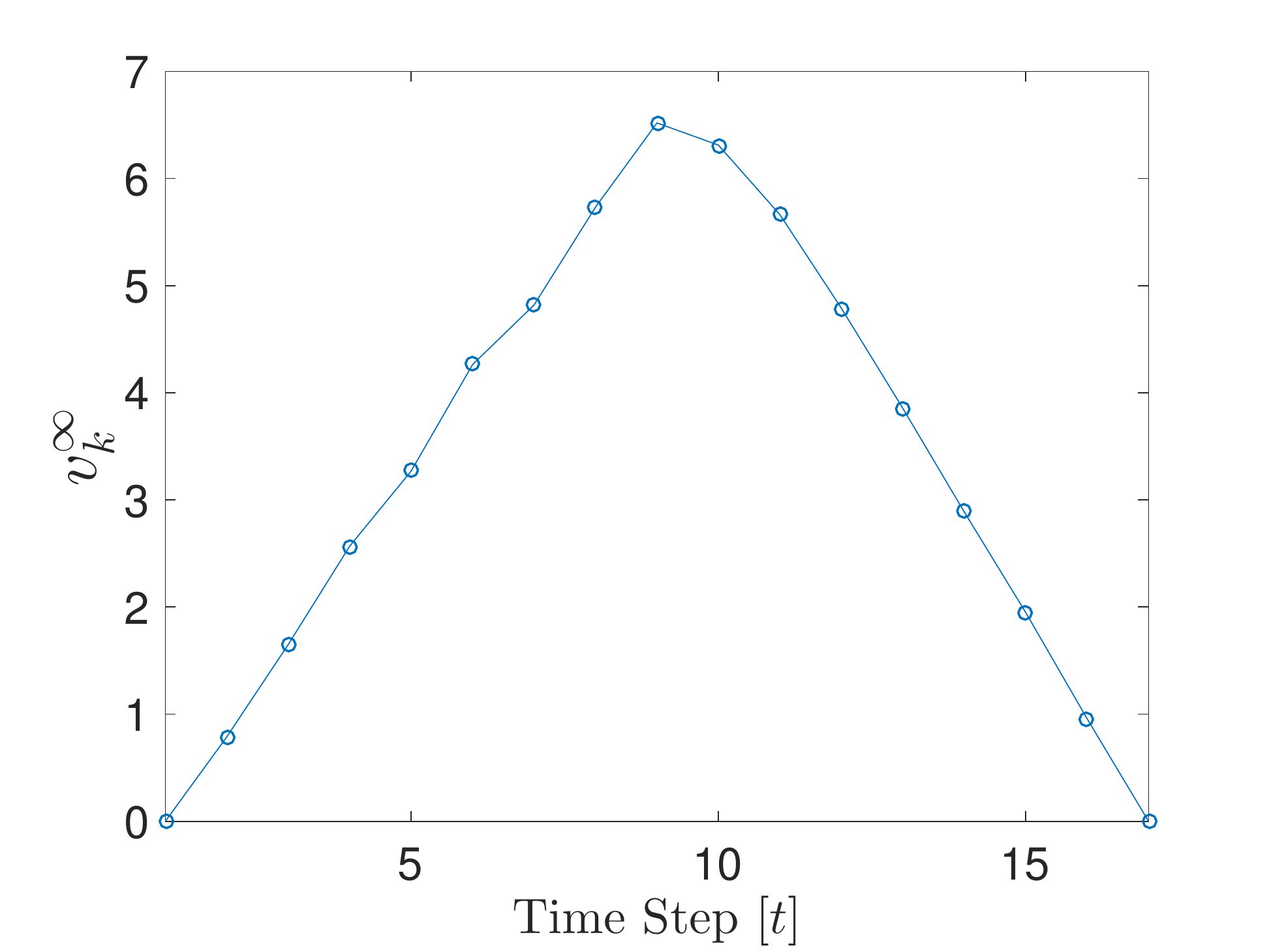}
	\caption{The velocity profile $v_k^{\infty}$ of the steady state trajectory ${\bf{x}^{\infty}}$.}
	\label{Res2:DubinsSteadyStateVelocity}
\end{figure}

After $4$ iterations the LMPC (\ref{eq:Constraints}) and (\ref{eq:MPC}) converges to the steady state solution shown in Figure \ref{Res2:DubinsSteadyState}. Table \ref{table:CostDubins} shows that the cost is decreasing until convergence is achieved. The steady state inputs are reported in Figure \ref{Res2:DubinsSteadyStateInputs}, we notice that the controller acceleration is very close to the boundary as we would expect from the optimal solution to the minimum time Problem (\ref{eq:Dubins}). In particular, the LMPC (\ref{eq:Constraints}) and (\ref{eq:MPC}), similarly to a bang-bang \cite{liberzon2012calculus} controller, accelerates until it reaches the midpoint between the initial and final position and afterwards it decelerates to reach the $x_F$ with zeros velocity, as shown in Figure \ref{Res2:DubinsSteadyStateVelocity}. Finally, we underline that in discrete time the minimum time cost is given by the number of time steps needed to reach the terminal point, therefore it is not surprising that the acceleration is not saturated all time steps. Indeed an acceleration profile similar to the one shown in Figure \ref{Res2:DubinsSteadyStateInputs} that saturates the acceleration at all time steps would lead to a trajectory with the same associated cost.

We performed an additional step to verify the (local) optimality of the steady state trajectory ${\bf{x}^{\infty}}$ at which the LMPC (\ref{eq:Constraints}) and (\ref{eq:MPC}) converged.
We solved problem  (\ref{eq:Dubins}) with an horizon $N = 16$ by using an interior point nonlinear solver \cite{c31} initialized with  the trajectory obtained with our proposed approach at steady-state. We confirmed that the locally optimal solution of the solver coincides with the steady state solution of the LMPC (\ref{eq:Constraints}) and (\ref{eq:MPC}).


\textbf{\begin{table}[h!]
		\caption{Cost of the LMCPC at each $j$-th iteration}\label{table:CostDubins}
		\centering
		\begin{tabular}{lll}
			\multicolumn{3}{l}{~~\textbf{Iteration}}{~~\textbf{Iteration Cost}}                   \\ \hline
			& $j = 0$    & ~~~~$39$ \\
			& $j =1$     & ~~~~$21$  \\
			& $j =2$     & ~~~~$18$  \\
			& $j =3$     & ~~~~$17$  \\
			& $j =4$     & ~~~~$16$  \\
			& $j =5$     & ~~~~$16$  \\
		\end{tabular}
	\end{table}}

\subsection{Dubins Car with Obstacle and Unknown Acceleration Saturation}
Consider the minimum time Dubins car problem (\ref{eq:Dubins}) presented in the previous example. We assume in this section that the saturation limit $s$ is unknown.
We use a sigmoid function $\frac{a_k}{\sqrt{1 + a_k^2}}$ as a continuously differentiable approximation of the
saturation function and reformulate (\ref{eq:Dubins}) as
\begin{subequations}\label{eq:SigDubins}
	\begin{align}
	&J_{0\rightarrow \infty}^*(x_S)=\min_{\begin{smallmatrix} \theta_0, \theta_1,\ldots\\ a_0,a_1,\ldots \end{smallmatrix}} \sum\limits_{k=0}^{\infty} \mathds{1}_k  \label{eq:SigDubins0}\\
	& {\textrm{s.t. }} \notag\\
	&x_k = \begin{bmatrix} z_{k+1} \\ y_{k+1} \\ v_{k+1} \end{bmatrix}= \begin{bmatrix} z_{k} \\ y_{k} \\ v_{k} \end{bmatrix} + \begin{bmatrix} v_k cos(\theta_k)\\ v_k sin(\theta_k)\\ s \frac{a_k}{\sqrt{1 + a_k^2}} \end{bmatrix},~\forall k\geq 0 \label{eq:SigDubins1}\\
	&x_0=x_S = [0~ ~ 0 ~~ 0]^T,\label{eq:SigDubins2}\\
	& \frac{(z_k - z_{obs})^2}{a_e^2} + \frac{(y_k - y_{obs})^2}{b_e^2} \geq 1, ~~ \forall k\geq 0.\label{eq:SigDubins5}
	\end{align}
\end{subequations}
where the indicator functino  $\mathds{1}_k$ is defined in (\ref{eq:Indicator}). The state vector $x_k = [z_k, y_k, v_k]$ collects the car position of car on the $Z$-$Y$ plane and the velocity, respectively. The inputs are the acceleration $a_k$ and the heading angle $\theta_k$. Finally, $s$ represents the unknown saturation limit. As in the previous example, we set $x_F = [54, 0, 0]^T$, $a_e=8$ and $b_e=6$. The vehicle model uses a saturation limit $s=1$. This is unknown to the controller.

We apply the proposed LMPC on an augmented system to simultaneously estimated the saturation coefficient and to steer the system (\ref{eq:SigDubins1}) to the terminal point $x_F$. In order to archive this,
we define a saturation coefficient estimate, $\hat{s}_k$, and an error estimate $e_k = s - \hat{s}_k$. The idea of augmenting the system with an estimator and a related error dynamics is standard in  adaptive control strategies \cite{bai2009adaptive} \cite{goodwin1986rapprochement}. The objective of the controller is a trade off between estimating the saturation coefficient and steering the system to the terminal point ${x}_F$. The LMPC solves at time $t$ of the $j$-th iteration the following problem,
\begin{subequations}\label{eq:DubinsLMPC}
	\begin{align}
	J_{0\rightarrow N}^{\scalebox{0.4}{LMPC},j}(&x_t^j)=\min_{\small{\begin{smallmatrix} \theta_0,\ldots,\theta_{N}\\ a_0,\ldots,a_{N} \\ \delta_0,\ldots,\theta_{N} \end{smallmatrix}}}  \Big[ \sum_{k=0}^{N-1} w_e e_k^2 + \bar{\mathds{1}}_k \Big] + Q^{j-1}(x_N) \label{eq:DubinsLMPC0}\\
	\textrm{s.t.}& \notag\\
	\hat{x}_{k+1} =& \hat{f}(\hat{x}_k, \hat{u}_k)  = \begin{bmatrix} \hat{z}_{k} \\ \hat{y}_{k} \\ \hat{v}_{k} \\ \hat{s}_{k} \\ e_k \end{bmatrix} + \begin{bmatrix}  \hat{v}_k cos(\theta_k) \\ \hat{v}_k sin(\theta_k) \\\hat{s}_{k+1} \frac{a_k}{\sqrt{1+a_k^2}} \\ \delta_k \\ -\delta_k \end{bmatrix}, ~\forall k\geq 0 \label{eq:DubinsLMPC1}\\
	&\small{\hat{x}_0=x_t^j},\label{eq:DubinsLMPC2}\\
	&\small{ \frac{(\hat{z}_k - z_{obs})^2}{a_e^2} + \frac{(\hat{y}_k - y_{obs})^2}{b_e^2} \geq 1, ~~ \forall k\geq 0,}\label{eq:DubinsLMPC5}\\
	&\small{x_N \in ~\mathcal{SS}^{j-1}}, \label{eq:DubinsLMPC6}
	\end{align}
\end{subequations}
where $N = 4$ and the weight on the error estimate $w_e = 10$. The indicator function $\bar{\mathds{1}}_k$ in (\ref{eq:DubinsLMPC0}) is defined as
\begin{equation}
\begin{aligned}
\bar{\mathds{1}}_k= \begin{cases} 1, & \mbox{if } \hat{x}_k \notin \mathcal{X}_F\\
0, & \mbox{if } \hat{x}_k \in \mathcal{X}_F \end{cases}.
\end{aligned}
\end{equation}
where
 \begin{equation} \label{eq:TerminalSet}
 \mathcal{X}_F = \Bigg\{  \bar{x} = \small\begin{bmatrix}
 \hat{z }\\
 \hat{y} \\
 \hat{v}\\
 \hat{s} \\
 e
 \end{bmatrix} \in \rr^5 : \small\begin{bmatrix}
 \hat{z} \\
 \hat{y} \\
 \hat{v} \\
 \end{bmatrix} = x_F, \hat{s} \in \rr, e = 0   \Bigg\}.
 \end{equation}
 $\hat{f}(\cdot, \cdot)$ in (\ref{eq:DubinsLMPC1}) represents the dynamics update of the augmented system and the state vector $\hat{x}_k = [\hat{z}_k, \hat{y}_k, \hat{v}_k, \hat{s}_k, e_k]$ collects the estimate position on the $Z$-$Y$ plane, the car's velocity, the saturation coefficient estimator and the estimator error, respectively. The input vector $\hat{u}_k = [a_k, \theta_k, \delta_k]$ collects the acceleration, the steering and the estimate difference between two consecutive time steps, respectively. Equation (\ref{eq:DubinsLMPC2}) represents the initial condition and (\ref{eq:DubinsLMPC5}) the obstacle avoidance constraint. Constraint (\ref{eq:DubinsLMPC6}) enforces the terminal state into the  $\mathcal{SS}^{j-1}$ defined in equation (\ref{eq:SS}). Finally, in (\ref{eq:DubinsLMPC}) we have used a simplified notation to equation (\ref{eq:Constraints}). \\
Let at time $t$ of the $j$-th iteration ${\bf{u}}^{*,j}_{t:t+N|t}$ be the optimal solution to (\ref{eq:DubinsLMPC}), then we apply the first element of ${\bf{u}}^{*,j}_{t:t+N|t}$ to the system in (\ref{eq:DubinsLMPC1}) \\
\begin{equation}\label{eq:MPCLogic}
\begin{aligned}
u_t^j = u_{t|t}^{*,j}.
\end{aligned}
\end{equation}

\begin{figure*}[h!]
	\includegraphics[width=\textwidth]{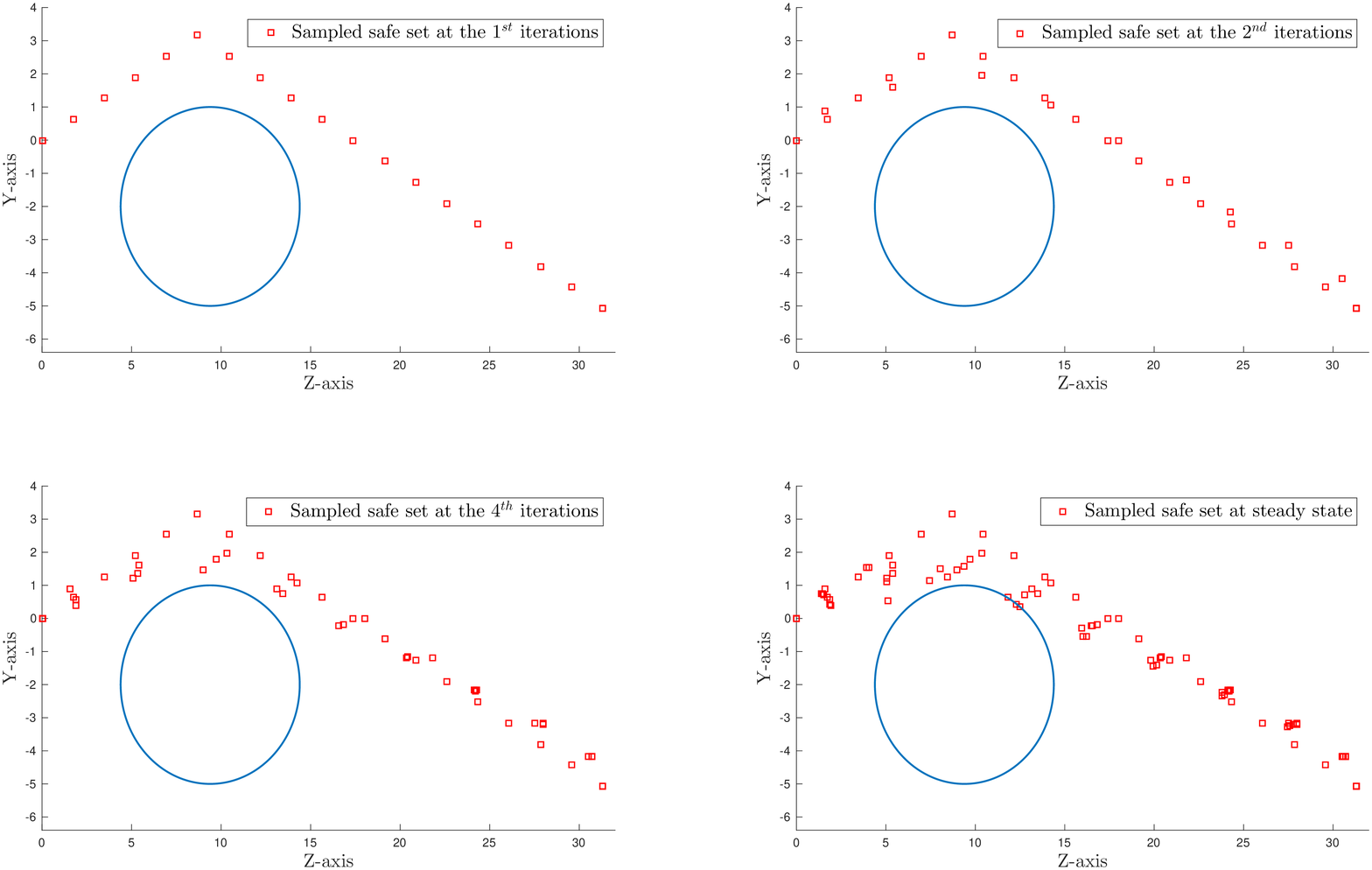}
	\caption{Sampled safe set evolution over the iterations.}
	\label{Res1:Learning}
\end{figure*}

\begin{figure}[h!]
	\centering
	\includegraphics[width=0.9\columnwidth]{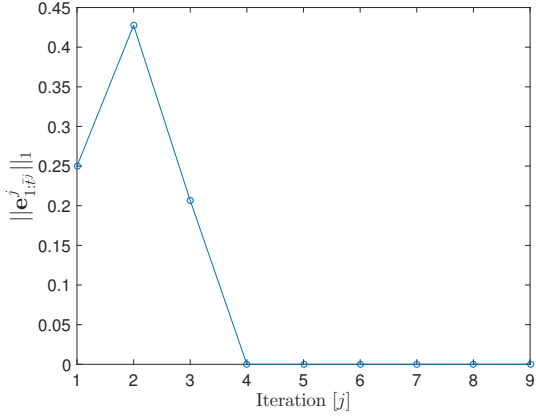}
	\caption{Evolution of the $1$-norm of the estimation error through the iterations.}
	\label{Res1:EstimationErrorVsIteration}
\end{figure}

\begin{figure}[h!]
	\centering
	\includegraphics[width=0.9\columnwidth]{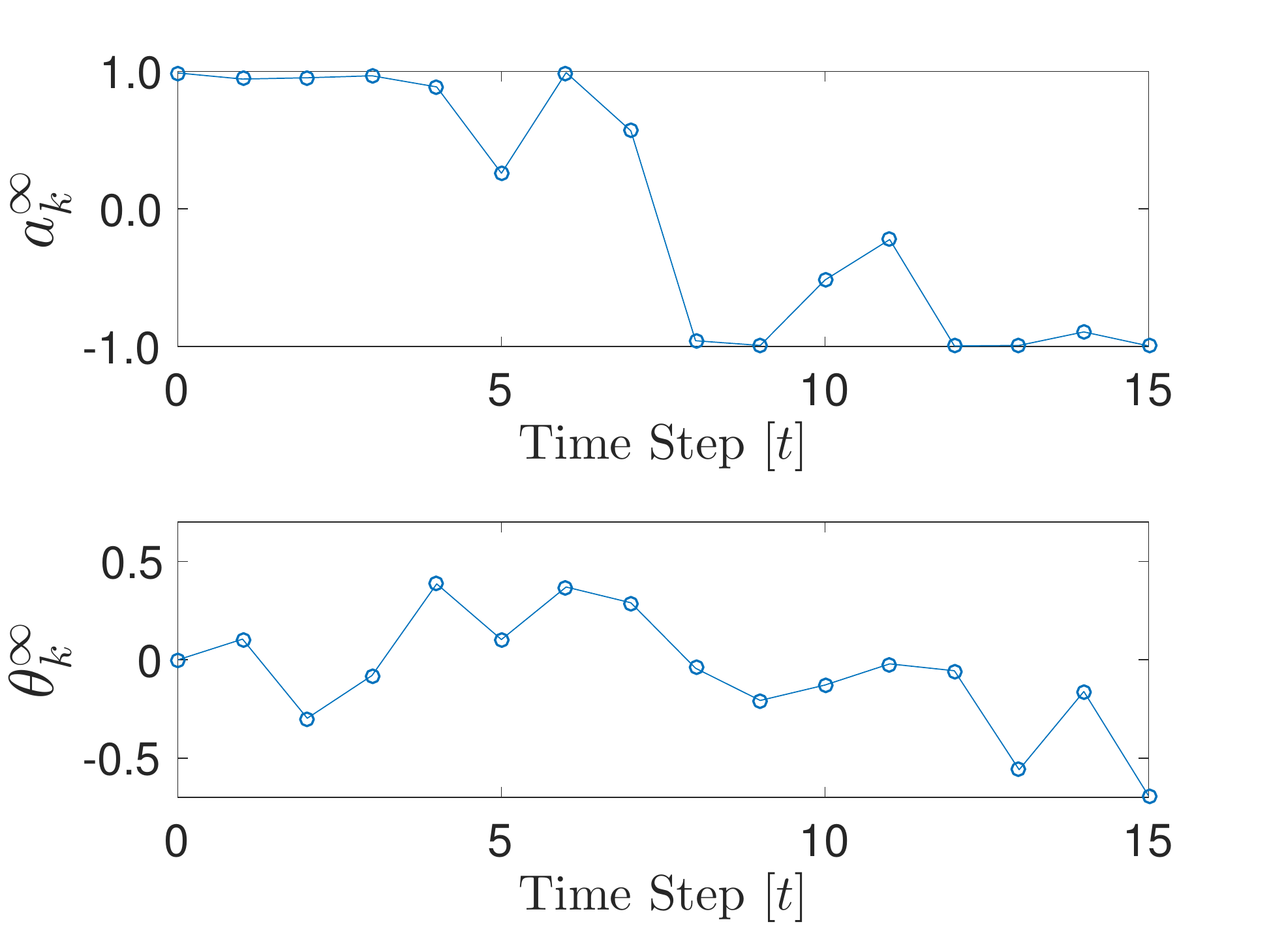}
	\caption{The acceleration $a_k^{\infty}$ and steering $\theta_k^{\infty}$ inputs associated with the steady state trajectory ${\bf{x}^{\infty}}$.}
	\label{Res1:Input}
\end{figure}

\begin{figure}[h!]
	\centering
	\includegraphics[width=0.9\columnwidth]{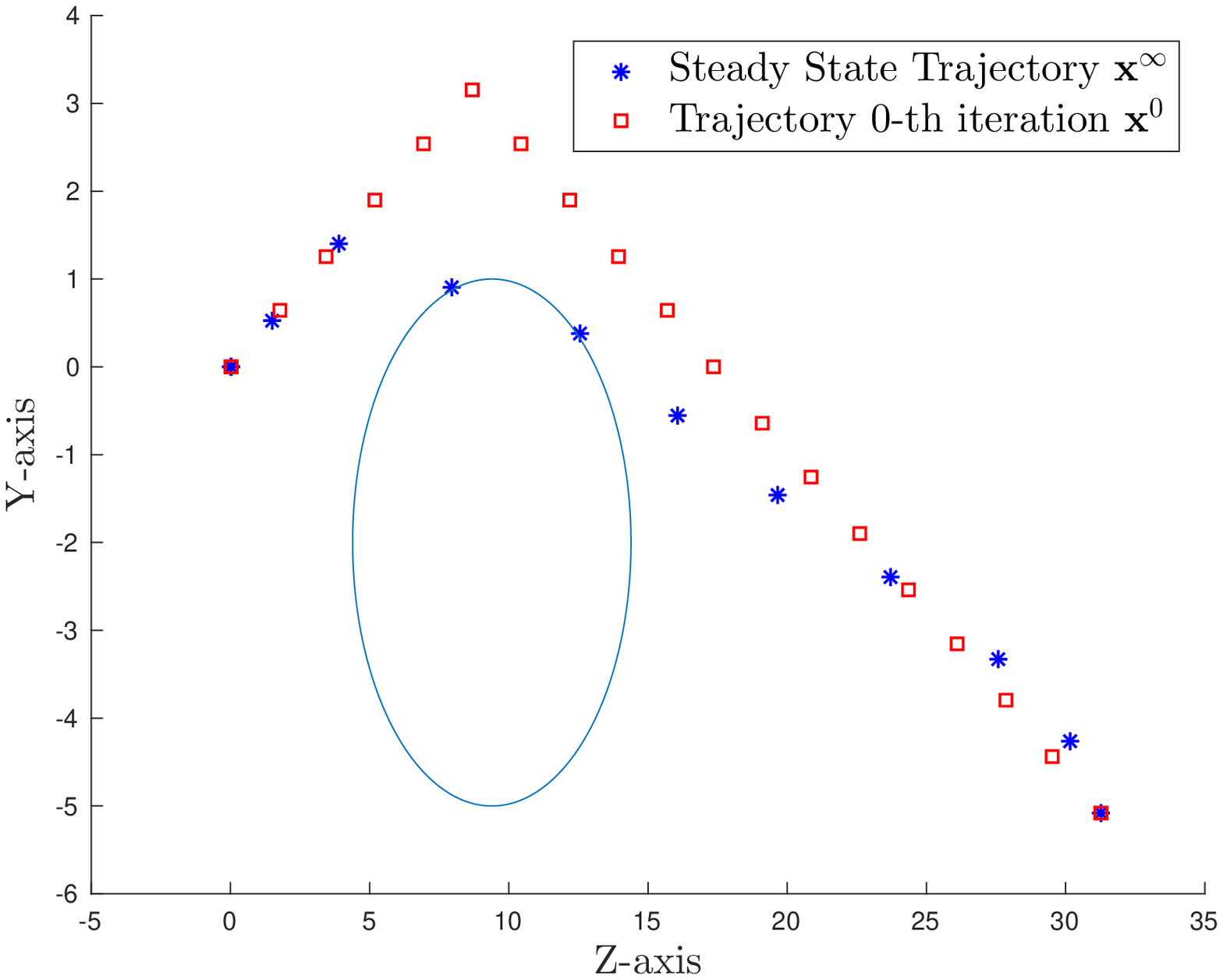}
	\caption{Steady state trajectory of the LMPC on the $Z-Y$ plane.}
	\label{Res1:TrajectoryDubisZY}
\end{figure}

We assume that at time $t$ of the $j$-th iteration the system state $x_t^j = [z_t^j, y_t^j, v_t^j]$ is measured and we estimate $e_t^j$ inverting the system dynamics (\ref{eq:SigDubins1}) and (\ref{eq:DubinsLMPC1})
\begin{equation}
e_{t}^j = \begin{cases}
\frac{y_{t}^j-\hat{y}_{t}^j - (y_{t-1}^j-\hat{y}_{t\text{-}1}^j)}{\frac{a_{t-1}^j}{\sqrt{1 + (a_{t-1}^j)^2}}}, & \mbox{If } \frac{a_{t-1}}{\sqrt{1 + a_{t-1}^2}} \neq 0\\
e_{t-1}^j & \mbox{otherwise}
\end{cases}.
\end{equation}\\
\begin{remark}\label{Remark:Optimality}
Consider a local optimal solution \mbox{${\bf{\bar{x}}^*} = [{\bf{z}}^*, ~{\bf{\hat{y}}}^*,~ {\bf{v}}^*,~ {\bf{\hat{s}}}^*,~ {\bf{e}}^*]^T$}
	 to problem (\ref{eq:DubinsLMPC}) defined over the infinite horizon. If $e_k^* = 0,~\forall k > 0$, (i.e., the algorithm has successfully identified the
friction saturation coefficient), then ${\bf{{x}}^*} =  [{\bf{z}}^*, ~{\bf{\hat{y}}}^*,~ {\bf{v}}^*]$ is a local optimal solution for the original problem (\ref{eq:Dubins}).
\end{remark}

Initialization of the LMPC (\ref{eq:DubinsLMPC}) is discussed in the Appendix.
For this example, Problem~(\ref{eq:Constraints}) can be reformulated as a Mixed-Integer Quadratic Program (MIQP). Further details on its solution can be found in  Section VII.A.2.


After $7$ iterations, the LMPC (\ref{eq:DubinsLMPC}), (\ref{eq:MPCLogic}) converges to a steady state solution. Figure \ref{Res1:Learning} illustrates the evolution of the sampled safe set through the iterations and Table \ref{table:CostLMPC} shows that the iteration cost is decreasing until convergence is reached.

Figure \ref{Res1:EstimationErrorVsIteration} shows the behavior of the $1$-norm of the error vector
\begin{equation}\label{eq:EstimtionError}
{\bf{e}}^j_{1:\infty} = [e_1^j, \dots, e_t^j, \dots].
\end{equation}
as a function of the iteration $j$. We notice that the LMPC (\ref{eq:DubinsLMPC}), (\ref{eq:MPCLogic}) correctly learns from the previous iterations decreasing the estimation error, until it identifies the unknown saturation coefficient (i.e. $e^{\infty}_k = 0~ \forall k > 0$).

The steady state inputs are reported in Figure \ref{Res1:Input}. One can observe that the LMPC (\ref{eq:DubinsLMPC}), (\ref{eq:MPCLogic}) saturates the acceleration constraints. The controller accelerates until it reaches the midpoint between the initial and final position and it decelerates afterwards, as we would expect from the optimal solution to a minimum time problem \cite{liberzon2012calculus}. Figure \ref{Res1:TrajectoryDubisZY} shows the steady state trajectory ${\bf{x}^{\infty}}$, and the feasible trajectory ${\bf{x}^{0}}$ at the $0$-th iteration. The LMPC (\ref{eq:DubinsLMPC}) and (\ref{eq:MPCLogic}) steers the system from the staring point $x_S$ to the final point $x_F$ in $16$ steps as the optimal solution to (\ref{eq:Dubins}) computed in the previous example.

\textbf{\begin{table}[h!]
		\caption{Optimal cost of the LMPC at each $j$-th iteration}\label{table:CostLMPC}
		\centering
		\begin{tabular}{llrl}
			\multicolumn{3}{l}{~~\textbf{Iteration}}{~~\textbf{Iteration Cost~~~~~~~~~~}}                   \\ \hline
			& $j = 0$    & ~~~$65.000000000000000$ \\
			& $j =1$     & ~~~$33.634529488066327$  \\
			& $j =2$     & ~~~~$24.216166714512450$  \\
			& $j =3$     & ~~~~$19.625000000001727$  \\
			& $j =4$     & ~~~~$19.625000000000004$  \\
			& $j =5$     & ~~~~$17.625000000022546$  \\
			& $j =6$     & ~~~~$17.625000000000000$  \\
			& $j =7$     & ~~~~$16.625000000000000$  \\
			& $j =8$     & ~~~~$16.625000000000000$  \\			\end{tabular}
	\end{table}}
		
\section{Practical Considerations}

\subsection{Computation}
The sampled safe set (\ref{eq:SS}) is a set of discrete points and therefore the terminal constraint in (\ref{eq:Constraints5}) is an integer constraint. Consequently, the proposed approach is computationally expensive also for linear system as the controller has to solve a mixed integer programming problem at each time step. In the following we discuss two different approaches to improve the computational burden associated with the proposed control logic.\\

\subsubsection{Convexifing the terminal constraint}

The computational burden associated with the finite time optimal control problem (\ref{eq:Constraints}) can be reduced relaxing the sampled safe to its convex hull,
and the $Q(\cdot)$ function to be its barycentric approximation. For more details on barycentric approximation we refer to \cite{c16}.  This relaxed problem is convex if the system dynamics is linear and the stage cost is convex. Furthermore,  for linear system and convex stage cost, the relaxed approach preserves the properties showed in \textit{Theorems 1-3} of \cite{LMPCLinear}. When the system is nonlinear, it is still possible to apply the convex relaxation but guarantees are, in general,  lost. In \cite{LMPCracing}, this relaxed approach has been successfully applied in real time to the nonlinear minimum time autonomous racing problem, where the LMPC is used to improve the vehicle's lap time over the iterations. A video of a more recent implementation on the Berkeley Autonomous Racing Car (BARC) platform can be found here: {\footnotesize{ \url{https://automatedcars.space/home/2016/12/22/learning-mpc-for-autonomous-racing} }}.  \\

\subsubsection{Parallelize Computations} The structure of the LMPC can be exploited to design an algorithm that: \emph{i}) use a subset of the sampled safe in the (\ref{eq:Constraints}), \emph{ii}) can be parallelized. In particular, one can compute an upper and lower bound to the optimal solution of problem (\ref{eq:Constraints}). These bounds allow to reduce the complexity of (\ref{eq:Constraints}) without loosing the guarantees proven in \textit{Theorems 1-3}. More details are discussed next.

First, we notice that at time $t,~ \forall t > 0$ it is possible to compute an upper bound on the optimal cost of problem (\ref{eq:Constraints}), using the solution computed at time $t-1$. In particular, from equations (\ref{eq:RunningCost}) and  (\ref{eq:LyapProof2}) we have,\\
\begin{equation}
J_{0\rightarrow N}^{\scalebox{0.4}{LMPC},j}(x_{t}^j)-J_{0\rightarrow N}^{\scalebox{0.4}{LMPC},j}(x_{t-1}^j) \leq - h(x_{t-1}^{j},u_{t-1}^{j}) \leq 0,
\end{equation}
which implies that at time $t$ an upper bound on the optimal cost is given by
\begin{equation} \label{eq:upperbound}
J_{0\rightarrow N}^{\scalebox{0.4}{LMPC},j}(x_{t}^j) \leq J_{0\rightarrow N}^{\scalebox{0.4}{LMPC},j}(x_{t-1}^j).
\end{equation}

In order to compute a lower bound, let (\ref{eq:OptimalSolutionMPC}) be the optimal solution to (\ref{eq:Constraints}), then at the $j$-th iteration\\
\begin{equation}
J_{t\rightarrow t+N}^{\scalebox{0.4}{LMPC},j}(x_t^j)= \sum_{k=t}^{t+N-1}  h(x_{k|t}^{*,j},u_{k|t}^{*,j}) + Q^{j-1}(x_{t+N|t}^{*,j}).\\
\end{equation}
As Problem (\ref{eq:DubinsLMPC}) is time-invariant and $h(\cdot,\cdot)$ is positive definite (\ref{eq:RunningCost}), we have
\begin{equation} \label{eq:lowebound}
J_{0\rightarrow N}^{\scalebox{0.4}{LMPC},j}(x_t^j) \geq Q^{j-1}(x_{t+N|t}^{*,j}), ~~ \forall x_{t+N|t}^{*,j} \in \mathcal{SS}^{j-1}.
\end{equation}
Combining the upper bound (\ref{eq:upperbound}) and the lower bound (\ref{eq:lowebound}), we obtain
\begin{equation}
Q^{j-1}(x_{t+N|t}^{*,j}) \leq J_{0\rightarrow N}^{\scalebox{0.4}{LMPC},j}(x_t^j) \leq J_{0\rightarrow N}^{\scalebox{0.4}{LMPC},j}(x_{t-1}^j).
\end{equation}
Therefore at optimum we have that
\begin{equation} \label{eq:upplow}
Q^{j-1}(x_{t+N|t}^{*,j}) \leq J_{0\rightarrow N}^{\scalebox{0.4}{LMPC},j}(x_{t-1}^j).
\end{equation}

Define $\mathcal{RS}_t^{j-1}$ as the set of points which satisfy condition (\ref{eq:upplow}),
\begin{equation}
\mathcal{RS}^{j-1}_{t} = \{ x \in \mathcal{SS}^{j-1}_t : Q^{j-1}(x) \leq J_{0\rightarrow N}^{\scalebox{0.4}{LMPC},j}(x_{t-1}^j) \},
\end{equation}
then, from equation (\ref{eq:upplow}), we deduce that for $t>0$
\begin{equation}
x_{t+N|t}^{*,j} \in \mathcal{RS}^{j-1}_{t}  \subseteq \mathcal{SS}^{j-1}.
\end{equation}
The set $\mathcal{RS}$ can be used in place of $\mathcal{SS}$ in order to reduce computational complexity.

The following {\bf{Algorithm 1}} uses this idea to solve the LMPC (\ref{eq:Constraints}), (\ref{eq:MPC}). {\bf{Algorithm 1}} was used for the Dubins Car example with the nonlinear solver Ipopt~\cite{c31}.

\begin{figure}[!h]
	\removelatexerror
	\begin{algorithm}[H]
		\SetAlgoLined
		Read measurements and update $x_t^j$ and $t$. \\
		\eIf{$t > 0$}{
			Compute $\mathcal{RS}^{j-1}_t$}
		{Set $\mathcal{RS}^{j-1}_t = \mathcal{SS}^{j-1}$}
		$n = 0$\\
		\For{\textit{all} $x \in  \mathcal{RS}^{j-1}_t$}{
			In (\ref{eq:Constraints}), set $\mathcal{SS}^{j-1} = x$\\
			Solve (\ref{eq:Constraints}) using a nonlinear optimization solver.\\
			Set $\bar{u}_n = u_{t|t}^{*,j}$ and $\bar{J}_n = J_{t\rightarrow t+N}^{\scalebox{0.4}{LMPC},j}(x_t^j)$\\
			$n = n +1$
		}
		Find $n^* = \arg \min_n \bar{J}_n$ \\
		Apply $u_t^j = \bar{u}_{n^*}$
		\caption{Compute $u_t^j$ at time $t$ of the $j$-th iteration}
	\end{algorithm}
\end{figure}


\subsection{Uncertainty}
The paper uses a deterministic framework and the theoretical guaranties have been demonstrated only for the deterministic case. This is the case of the vast majority of seminal papers on MPC~\cite{c12, zheng1995stability, garcia1989model, de2000contractive}.
In the presence of disturbances, as for all deterministic MPC schemes,  all the guarantees are lost.
However, one can build on the proposed results to formulate a \textit{stochastic iterative learning MPC}.
For instance if disturbance is modeled as a Gaussian process the chance constraint can be converted to  deterministic second order cone constraint \cite{calafiore2007linear}, which can be handled with the proposed control logic. Furthermore, the proposed control logic can be extended to a \textit{robust iterative learning MPC} when the disturbance is bounded and the system is linear. Under these assumptions the robust MPC can be formulated in a deterministic control problem tightening the constraints \cite{kouvaritakis2015model}. In particular, the robust MPC can be designed on a nominal model where the tightening of the state constraints is computed to guarantee that the original system satisfies the nominal constraints for all the disturbance values \cite{kouvaritakis2015model}.
This is topic of further  investigation.

\section{Conclusions}
In this paper, a reference-free learning nonlinear model predictive control that exploits information from the previous iterations to improve the performance of the closed loop system over iterations is presented. A safe set and a terminal cost, learnt from previous iterations, allow to guarantee the recursive feasibility and stability of the closed loop system. Moreover, we showed that if the closed-loop system converges to steady state trajectory then this trajectory is locally optimal for an approximation of the infinite horizon control problem. We tested the proposed control logic on an infinite horizon linear quadratic regulator with constraints (CLQR) to shown that the proposed control logic converges to the optimal solution of the infinite optimal control problem. Finally, we tested the control logic on nonlinear minimum time problem optimal control problem and we showed that the properties of the proposed LMPC can be used to simultaneously estimate unknown system parameters and to generate a state trajectory that pushes system performance.

\section{Acknowledgments}
We thank the reviews for their feedback on the manuscript.

\section{Appendix}
In order to compute a feasible trajectory that steers system (\ref{eq:DubinsLMPC1}) from the initial state $\bar{x}_0 = [x_0,\hat{s}_0, e_0]^T$ into $\mathcal{X}_F$ we
used a greedy approach described next.
First, we set $\delta_k = 0, ~\forall k = 1,\ldots, N-1$. Therefore, from (\ref{eq:DubinsLMPC1}), we have that
\begin{subequations}
	\begin{align}
	\hat{s}_k &= \hat{s}_0 , ~\forall k = 1,\ldots, N-1\\
	e_k &= e_0 , ~\forall k = 1,\ldots, N-1.
	\end{align}
\end{subequations}

Afterwards, we selected an initial guess for the saturation coefficient estimate $\hat{s}_0 = 0.25$  and given the following input structure
\begin{subequations}\label{eq:Input2}
	\begin{align}
	\theta_k &= \tilde{\theta},~~~~~~~\forall k = 1, \ldots, N_s \\
	\theta_k &= -\tilde{\theta}, ~~~~~\forall k = N_s+1, \ldots, N-1 \\
	a_k &= \tilde{a}, ~~~~~~~\forall k = 1, \ldots, \bar{N}_s\\
	a_k &=0, ~~~~~~~\forall k = \bar{N}_s+1, \ldots, N-\bar{N}_s \\
	a_{N-1} &= -\tilde{a}, ~~~~~\forall k = N-\bar{N}_s+1, \ldots, N
	\end{align}
\end{subequations}
we generated a set of trajectories using different sets of parameters $\tilde{\theta}, N_s, \bar{N_s}, \tilde{a}, N$. Among the generated trajectories, we used the one
minimizing the following quantity
\begin{equation}
||\small\begin{bmatrix}
z_{N-1} \\
\bar{y}_{N-1} \\
v_{N-1} \\
\hat{s}_{N-1} \\
e_{N-1}\end{bmatrix}  -
\small\begin{bmatrix}
x_F \\
\hat{s}_{N-1} \\
e_{N-1}		
\end{bmatrix}||_2^2
\end{equation}
to warm-start a nonlinear optimization problem which allowed us to find the following $N-1$ step trajectory
\begin{equation} \label{eq:EXII}
{\bf{\bar{x}}}^0_{0:N-1} = \Big[\bar{x}_0^0, \ldots, \bar{x}_{N-1}^0 = \begin{bmatrix}
x_F \\
\hat{s}_{N-1} \\
e_{N-1}		
\end{bmatrix}\Big],
\end{equation}
and the related input sequence
\begin{equation} \label{eq:InputsSeq}
(\theta_k^0, a_k^0), ~ \forall k = 1,\ldots, N-1.
\end{equation}
Afterwards the input sequence (\ref{eq:InputsSeq}) are applied to the system (\ref{eq:Dubins1}) to compute
\begin{equation}
{\bf{{x}}}^0_{0:N-1} = [{x}_0^0, \ldots, {x}_{N-1}^0].
\end{equation}
Then realized trajectories  ${\bf{\bar{x}}}^0_{0:N-1}$ and ${\bf{x}}^0_{0:N-1}$ are used to compute the error, which from equations (\ref{eq:Dubins1}) and (\ref{eq:DubinsLMPC1}), is given by
\begin{equation}
e_{k+1} = \begin{cases}
\frac{y_{k+1}-\hat{y}_{k+1} - (y_{k}-\hat{y}_{k})}{\frac{a_k}{\sqrt{1 + a_k^2}}}, & \mbox{If } \frac{a_k}{\sqrt{1 + a_k^2}} \neq 0\\
e_k & \mbox{else}
\end{cases}
\end{equation}
$\forall k = 0, \ldots, N-2 $.\\
Finally, we selected
\begin{subequations} \label{eq:Input3}
	\begin{align}
	\theta_N^0 &= a_N^0 = 0 \\
	\delta_N^0 &= e_{N-1}
	\end{align}
\end{subequations}
to regulate $e_{N-1}^0$ to zero steering $\bar{x}_{N-1}^0$ into $\mathcal{X}_F$. Concluding, the $N$ steps trajectory which extends the trajectory in (\ref{eq:EXII}) using (\ref{eq:Input3}),
\begin{equation}
{\bf{\bar{x}}}^0_{0:N} = \Big[\bar{x}_0^0, \ldots, \bar{x}_{N}^0 = \begin{bmatrix}
x_F \\
\hat{s}_{N} \\
0		
\end{bmatrix}\Big]
\end{equation}
steers system (\ref{eq:DubinsLMPC1}) into $\mathcal{X}_F$ and it can be used to build $\mathcal{SS}^0$ and $Q^0(\cdot)$.

\bibliographystyle{IEEEtran}
\bibliography{IEEEabrv,mybibfile}

\begin{biography}[{\includegraphics[width=1in,height=1.25in,clip,keepaspectratio]{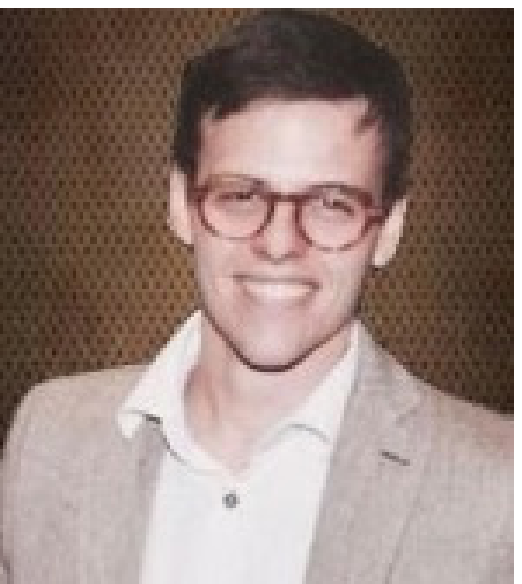}}]{Ugo Rosolia}
received the B.S. and M.S. (\textit{cum laude}) degrees in mechanical engineering from the Politecnico di Milano, Milan, Italy, in 2012 and 2014, respectively. He is currently pursuing the Ph.D. degree in mechanical engineering with the University of California at Berkeley, Berkeley, CA, USA.

He was a Visiting Scholar with Tongji University, Shanghai, China, for the Double Degree Program PoliTong from 2010 to 2011. During his M.S. degree, he was a Visiting Student for two semesters with the University of Illinois at Urbana${\text-}$Champaign, Urbana, IL, USA, sponsored by a Global E3 Scholarship. He was a Research Engineer with Siemens PLM Software, Leuven, Belgium, in 2015, where he was involved in the optimal control algorithms. His current research interests include nonlinear optimal control for the centralized and decentralized system, the iterative learning control, and the predictive control.
\end{biography}

\vspace{-0.8cm}
{\tiny }
\begin{biography}[{\includegraphics[width=1in,height=1.25in,clip,keepaspectratio]{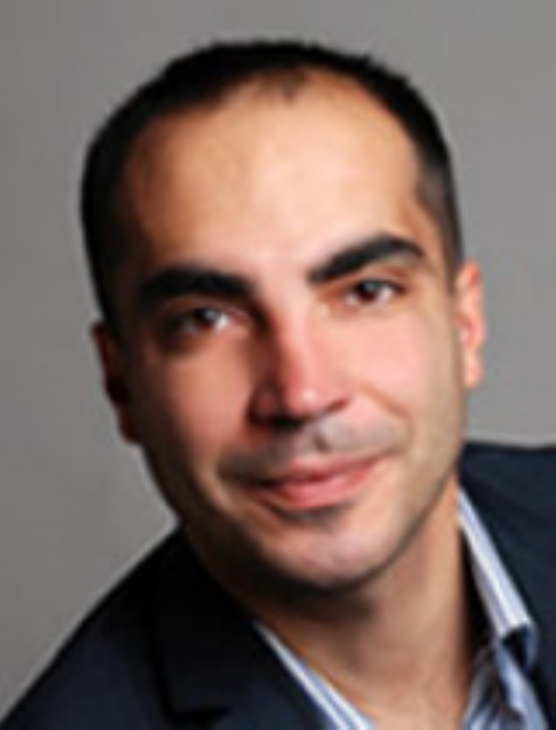}}]{Francesco Borrelli}
received the Laurea degree in computer science engineering from the University of Naples Federico II, Naples, Italy, in 1998, and the Ph.D. degree from ETH-Zurich, Zurich, Switzerland, in 2002.

He is currently an Associate Professor with the Department of Mechanical Engineering, University of California, Berkeley, CA, USA. He is the author of more than 100 publications in the field of predictive control and author of the book \textit{Constrained Optimal Control} of Linear and Hybrid Systems
(Springer Verlag). His research interests include constrained optimal control, model predictive control and its application to advanced automotive control and energy efficient building operation.

Dr. Borrelli was the recipient of the 2009 National Science Foundation CAREER Award and the 2012 IEEE Control System Technology Award. In 2008, he was appointed the chair of the IEEE technical committee on automotive control.
\end{biography}\vfill

\end{document}